\documentclass[12pt,thmsa,a4paper]{article}
\usepackage{sw20elba}

\input tcilatex
\QQQ{Language}{
British English
}

\begin{document}

\title{\textbf{Principles of Discrete Time Mechanics:}\\
\textbf{\ I. Particle Systems}}
\author{George Jaroszkiewicz and Keith Norton \\
Department of Mathematics, University of Nottingham\\
University Park, Nottingham NG7 2RD, UK}
\date{19$^{th}$ December 1996}
\maketitle

\begin{abstract}
\textit{We discuss the principles to be used in the construction of discrete
time classical and quantum mechanics as applied to point particle systems.
In the classical theory this includes the concept of virtual path and the
construction of system functions from classical Lagrangians, Cadzow's
variational principle applied to the action sum, Maeda-Noether and Logan
invariants of the motion, elliptic and hyperbolic harmonic oscillator
behaviour, gauge invariant electrodynamics and charge conservation, and the
Grassmannian oscillator. First quantised discrete time mechanics is
discussed via the concept of system amplitude, which permits the
construction of all quantities of interest such as commutators and
scattering amplitudes. We discuss stroboscopic quantum mechanics, or the
construction of discrete time quantum theory from continuous time quantum
theory and show how this works in detail for the free Newtonian particle. We
conclude with an application of the Schwinger action principle to the
important case of the quantised discrete time inhomogeneous oscillator. }
\end{abstract}

\section{Introduction}

THERE are various circumstances in mechanics where it is convenient or
necessary to replace the continuous time (temporal evolution) parameter with
a discrete parameter. Computer simulation of waves is an example where the
configuration of a system at time $t+T$ is calculated from a knowledge of
its configuration at times $t$ and $t-T$. There have been various attempts
to construct classical and quantum mechanical theories based on this notion,
such as the work of Cadzow \cite{CADZOW.70}, Logan \cite{LOGAN.73}, Maeda 
\cite{MAEDA.81} and Lee \cite{LEE.83}. The work of Yamamoto et al. \cite
{YAMAMOTO.95A,YAMAMOTO.95B} and Klimek \cite{KLIMEK.93} indicates that the
subject continues to receive attention.

This paper considers the question: by which principles if any should
continuous time mechanical theories be discretised, that is, turned into
discrete time analogues? By discretisation we do not mean the numerical
approximation of continuous time mechanics such as the work of Bender et al. 
\cite{BENDER.85A}. Neither do we discretise space or the dynamical degrees
of freedom. Our attention is fixed solely on replacing a continuous
dynamical evolution parameter with a discrete parameter. In this and the
following paper, \textit{Paper II} on discrete time classical field theory 
\cite{J&N-II}, our interest is the construction of exact, self consistent
discrete time mechanics with well specified principles, equations of motion
and predictions. This is motivated by the notion that at some unimaginably
small scale, time is really discrete. This has echoes in modern theories
such as string theory and quantum gravity, where the Planck time of $%
10^{-43}~$seconds sets a scale at which conventional notions of space and
time break down.

It could be argued that relativity requires a symmetrical treatment between
time and space but this leads to the situation of a space-time lattice
approach which has lost all relativistic symmetries and rotational
invariance. We argue that relativity does distinguish between timelike and
spacelike, and by discretising only time our approach reduces the break with
relativity to a minimum. Lorentz covariance is broken in our approach to
field theories, discussed in \textit{Paper II}, but the residual Euclidean
invariances permit the construct of particle like states.

It may be felt objectionable that there is no natural concept of velocity in
discrete time mechanics. It could be argued that this lack destroys our
intuitive feeling for dynamics based on the notion of (say) a particle
system evolving from an initial position and an initial velocity. The right
way to see the situation is in terms of real numbers. In continuous time
mechanics, we normally consider a particle as having an instantaneous
position and an instantaneous velocity (we exclude Brownian particle
dynamics from our definition of continuous time mechanics). This information
requires two real numbers for every degree of freedom. In discrete time
mechanics, there is no natural concept of simultaneity analogous to this.
What we mean by a ''particle'' is something with a position at time $t$ 
\textit{and} a position at time $t-T$, which also requires two real numbers
for every degree of freedom. A particle here is more properly associated
with the \textit{link} between two successive points in discrete time,
rather than those times separately. So ultimately, the only major difference
in principle between continuous time and discrete time mechanics is the lack
of the limit process $T\rightarrow 0$.

One problem with discrete time mechanics is a lack of guiding principles at
key places, which our series of papers attempts to address. For example,
consider the discretisation of a system with Lagrangian $L=\tfrac 12m{\dot{x}%
}^2-V(x)$. The approach taken by most authors would be to replace the
temporal derivatives by differences, symmetrise the potential in some way,
derive the analogue of the Euler-Lagrange equation, and finally evolve the
system according to the resulting difference equation. Quantities such as
the energy $E=\frac 12m\dot{x}^2+V(x)$ which are conserved in continuous
time mechanics would be monitored by calculating the value $E_D$ of the
discretised Hamiltonian.

It is more than likely however that a naive discretisation of the
Hamiltonian would result in an expression $E_D^{\prime }$ which is not
exactly conserved. This has been discovered by many authors. It is a
particular merit of Lee's approach \cite{LEE.83} that an invariant analogous
to the energy drops out of the formalism, but only at the expense of a
dynamically evolving discrete time interval.

It is somewhat surprising therefore that a computer simulation based on the
above principles should be judged good or bad according to how constant $%
E_D^{\prime }$ remains. In the absence of any proper principle for the
construction of invariants of the motion it should come as no surprise to
find that occasionally a quantity such as $E_D^{\prime }$ will vary
enormously and unpredictably during the course of a simulation. This happens
because there are really three systems being confused; 1) the original
continuous time theory, 2) a discrete time system evolving exactly according
to some well defined discretised Euler-Lagrange equation, and 3) some
unknown discrete time system for which the naively discretised energy $%
E_D^{\prime }$ would be an exact constant of the motion, but only for
evolution under its own discretised equation of motion, which could be very
different to the equation of motion for 2). On top of this there may be
numerical uncertainties induced by the computer algorithms used. Seen in
this light, it would seem a wise policy to discretise according to definite
principles which would establish conserved quantities rigorously. The
construction of invariants of the motion therefore becomes one of the
principal objectives of discrete time mechanics.

An important first step in the process of constructing rigorous discrete
time mechanics was the introduction of a discrete time action principle.
This was done by Cadzow \cite{CADZOW.70}, giving a discrete time analogue of
the Euler-Lagrange equation. We shall call such an equation a \textit{%
Cadzow's} \textit{equation} for the system. The construction of constants of
the motion was considered by Maeda \cite{MAEDA.81} in the case of continuous
symmetries, whilst the construction of constants of the motion analogous to
the energy had been considered earlier by Logan \cite{LOGAN.73}.

Various features found in continuous time mechanics have discrete time
analogues, including Noether's theorem, whilst certain other features either
do not or cannot have discrete time analogues. An particular problem arises
for example with Hamiltonian evolution and equations of motion derived using
Poisson brackets. Not only is there no possibility here of an infinitesimal
translation in time (which thereby renders the notion of a Hamiltonian
problematical) but there is no natural concept of velocity as a limit
either. This makes the standard definition of conjugate momentum as the
partial derivative of the Lagrangian with respect to a velocity just as
problematical. This has not prevented a number of authors from constructing
discrete time analogues of Poisson brackets, however, with various degrees
of success and utility, usually with the observation that the generator of
time translations is not conserved.

A feature of our approach is that we have found a clear principle for the
definition of conjugate momentum in discrete time mechanics. It turns out 
\textit{not} to be the partial derivative of the ''Lagrangian'' with respect
to a difference in general, but does reduce to it in various important
cases. In addition, we have avoided trying to construct Hamiltonians and
equations of motion derived via Poisson brackets. In our formalism the
Hamiltonian is displaced by a Logan invariant, if such a quantity can be
found. Fortunately such an object does exist for the important case of the
harmonic oscillator, which has ramifications in discrete time field theory
discussed in the next paper in this series.

The overall plan for this and subsequent papers is as follows. In this
paper, \textit{Paper I}, we restrict our attention to classical and quantum
point particle dynamics, reserving classical field theory to \textit{Paper II%
}, quantum field theory to \textit{Paper III}, and quantum electrodynamics
to \textit{Paper IV}.

Topics covered in \textit{Paper I} are as follows. First we introduce the
central concept of \textit{system function}. This replaces the Lagrangian as
the key to the dynamics. With the system function we can calculate equations
of motion, constructs invariants of the motion, and quantise the system. We
give a prescription for constructing the system function from a given
Lagrangian. We may use this prescription to embed into the system function
symmetries such as gauge invariance and hence construct electrodynamics.
Then we discuss the construction of invariants based on the work of Maeda,
Noether, and Logan, and apply it to the harmonic oscillator, which we
discuss in detail. This key system lies at the heart of particle field
theory, discussed in the following papers, and it displays some important
properties, such as a natural cutoff for particle energy for example. We
also discuss particle electrodynamics and the Grassmannian oscillator.

A major part of our programme is to develop discrete time quantum mechanics
and so we conclude our paper with a discussion of the principles for first
quantisation. This includes the concept of system amplitude, the
construction of unequal-time commutators, and compatible operators. We then
discuss the construction of discrete time quantum mechanics from standard
quantum mechanics via a stroboscopic approach and give an explicit example.
Finally we apply the Schwinger action principle to the discrete time
inhomogeneous harmonic oscillator to construct the Feynman propagator for
the oscillator, in anticipation of its use in field theory.

\section{Action integrals and action sums}

In continuous time mechanics Lagrangian dynamics is conventionally
formulated via an action principle based on the action integral 
\begin{equation}
A_{if}[\Gamma ]=\int_{t_i}^{t_f}dt\,L(\mathbf{q},\mathbf{\dot{q}},t),
\label{action integral}
\end{equation}
where $t_i$ and $t_f$ are the initial and final times respectively along
some given path $\Gamma $. In our version of discrete time mechanics we
postulate that the dynamical variables $\mathbf{q}(t)$ are observed or
sampled at a finite number of times $t_n,n=0,1,...,N$, where $t_0=t_i$ and $%
t_N=t_f$, such that the intervals $t_{n+1}-t_n$ are all equal to some
fundamental interval $T$. For convenience we will write $\mathbf{q}_n\equiv 
\mathbf{q}(t_n)$.

It is possible to develop a theory where the time intervals vary dynamically
along the path. Such a mechanics was considered by Lee \cite{LEE.83}. The
extension of our methods to that particular situation is left for a further
article.

In our formulation of discrete time mechanics we replace the action integral
(\ref{action
integral}) by an action sum of the form 
\begin{equation}
A^N[\Gamma ]=\sum_{n=0}^{N-1}F^n,  \label{sum}
\end{equation}
where $F^n\equiv F(\mathbf{q}_n,\mathbf{q}_{n+1},n)$ will be referred to as
the \textit{system function}. The system function has the same central role
in discrete time mechanics as the Lagrangian has in continuous time
mechanics. With it we may construct the equations of motion, define
conjugate momenta, construct constants of the motion and attempt to quantise
the system. In principle we could consider higher order system functions
which depended on (say) $\mathbf{q}_n$, $\mathbf{q}_{n+1},$..., $\mathbf{q}%
_{n+r},$ $r\geq $ $2,$ but the case $r=1$ represents the simplest
possibility which could give rise to non-trivial dynamics and will be
considered exclusively from now on. Such system functions are the discrete
time analogues of Lagrangians of the canonical form $L=L\left( \mathbf{q},%
\mathbf{\dot{q}},t\right) .$

Another reason for considering only a second order formulation $(r=1)$ is
its direct relationship to Hamilton's principal function, discussed
presently. Cadzow \cite{CADZOW.70} applied a variational principle to an
action sum such as (\ref{sum}) and derived the equation of motion 
\begin{equation}
\frac \partial {\partial \mathbf{q}_n}\left\{ F^{n-1}+F^n\right\} 
\stackunder{c}{=}{}~\mathbf{0},~~~~~0<n<N,  \label{Cadzow}
\end{equation}
where the symbol $\stackunder{c}{=}$ denotes an equality holding over a true
or dynamical trajectory. We shall refer to $\left( \ref{Cadzow}\right) $ as
a \textit{Cadzow's equation of motion} for the system. We now discuss the
interpretation of this equation.

Suppose we have a continuous time action integral of the form (\ref
{action
integral}). First, partition the time interval $\left[
t_0,t_N\right] $ into $N$ equal subintervals. Then the action integral may
be written as a sum of sub-integrals, i.e., 
\begin{equation}
A_{if}[\Gamma ]=\sum_{n=0}^{N-1}\int_{t_n}^{t_{n+1}}dt\,L(\mathbf{q}(t),%
\mathbf{\dot{q}}(t),t).
\end{equation}
Now suppose that we fixed the co-ordinates $\mathbf{q}_n$ at the various
times $t_0,t_1,...,t_N$ and then chose the path connecting each pair of
points ({$\mathbf{q}_n,\mathbf{q}_{n+1})$} to be the true or dynamical path,
that is, a solution to the Euler-Lagrange equations of motion for those
boundary conditions. If this partially extremised path is denoted by $\tilde{%
\Gamma}_c$ then we may write 
\begin{equation}
A_{if}[\tilde{\Gamma}_c]=\sum_{n=0}^{N-1}S^n,  \label{action sum}
\end{equation}
where $S^n\equiv S(\mathbf{q}_{n+1},t_{n+1};\mathbf{q}_n,t_n)$ is known as
Hamilton's principal function, being just the integral of the Lagrangian
along the true path from $\mathbf{q}_n$ at time $t_n$ to $\mathbf{q}_{n+1}$
at time $t_{n+1}$.

We recall now that the canonical momenta $\mathbf{p}_n^{(+)}$ , $\mathbf{p}%
_{n+1}^{(-)}$ at the end points {$\mathbf{q}_n,\mathbf{q}_{n+1}$} may be
obtained from Hamilton's principal function via the rule 
\begin{equation}
\mathbf{p}_{n+1}^{(-)}\equiv \frac \partial {\partial \mathbf{q}%
_{n+1}}S^n,\;\;\;\;\;\mathbf{p}_n^{(+)}\equiv -\frac \partial {\partial 
\mathbf{q}_n}S^n,
\end{equation}
where the superscript $(+)$ denote that the momentum at the initial time $t_i
$ carries information forwards\emph{,} whereas the superscript $(-)$ denotes
that the momentum at the final time $t_f$ is influenced by earlier dynamics
with respect to the temporal interval concerned. At this stage the action
sum (\ref{action sum}) has not been extremised fully, as the intermediate
points $\mathbf{q}_n,0<n<N$ have been held fixed.

Now suppose we went further and extremised (\ref{action sum}) fully by
variation of the previously fixed intermediate co-ordinates $\mathbf{q}%
_n,n=1,2,...,N-1$. Then we would find that 
\begin{equation}
\frac \partial {\partial \mathbf{q}_n}\left\{ S^{n-1}+S^n\right\} 
\stackunder{c}{=}\mathbf{0},\;\;\;\;\;0<n<N.  \label{formal}
\end{equation}
This equation may be understood as the condition that the canonical momentum
along the true path from $\mathbf{q}_i$ to $\mathbf{q}_f$ is continuous,
that is, $\mathbf{p}_n^{(+)}\stackunder{c}{=}\mathbf{p}_n^{(-)}$. We notice
immediately that (\ref{formal}) has the same formal structure as Cadzow's
equation (\ref{Cadzow}) provided we make the identification $%
F^n\leftrightarrow S^n$.

Another interpretation of Cadzow's equation is that it endows the action sum
with the additivity property of action integrals, which satisfy the
relations 
\begin{equation}
\int\limits_{t_0}^{t_1}dtL+\int\limits_{t_1}^{t_2}dtL=\int%
\limits_{t_0}^{t_2}dtL,\;\;\;\;t_0<t_1<t_{2.}
\end{equation}
This property holds for all trajectories in continuous time mechanics, not
just for the true or classical trajectory. In the case of system functions
we may write 
\begin{equation}
F(\mathbf{q}_{n-1},\mathbf{q}_n)+F(\mathbf{q}_n,\mathbf{q}_{n+1})\stackunder{%
c}{=}f(\mathbf{q}_{n-1},\mathbf{q}_{n+1})
\end{equation}
for some function $f$ of $\mathbf{q}_{n-1}$ and $\mathbf{q}_{n+1}$, because
Cadzow's equation (\ref{Cadzow}) is equivalent to the statement that $%
F^n+F^{n-1}$ is independent of \thinspace $\mathbf{q}_n$ along dynamical
trajectories. However, unlike action integrals, this property will not hold
off the true or classical trajectory in general.\ 

\section{System functions from Lagrangians}

$~~~$Two important ideas emerge from the similarity between (\ref{Cadzow})
and (\ref{formal}):

$\mathbf{i)}$ Although the concept of velocity as a limit does not occur in
discrete time mechanics, we will define a unique discrete time momentum $%
\mathbf{p}_n$ conjugate to $\mathbf{q}_n$ by the rule 
\begin{equation}
\mathbf{p}_n\equiv -\frac \partial {\partial \mathbf{q}_n}F^n.  \label{pmom}
\end{equation}
This should be compared with the approach of Yamamoto et al. \cite
{YAMAMOTO.95A,YAMAMOTO.95B} and most other workers, where the momentum is
defined as a derivative of a discretised Lagrangian with respect to a
difference. In our terms Cadzow's equation reduces simply to the statement
that we may also calculate this momentum via the rule 
\begin{equation}
\mathbf{p}_n\equiv \frac \partial {\partial \mathbf{q}_n}F^{n-1}.
\end{equation}

$\mathbf{ii)}$ We will construct a system function $F^n$ from the temporal
integral from $t_n$ to $t_{n+1}$ of a continuous time Lagrangian, the
question being which path to take. We cannot in general consider using the
true continuous time path, as this is meaningless in the context of discrete
time mechanics and normally not known to us. For the particularly important
case of the harmonic oscillator, however, we can evaluate Hamilton's
principal function precisely and this provides us with an important check on
our formalism. The path chosen in the construction of the system function
will be referred to as a \textit{virtual path}.

It is possible to choose from a number of possible virtual paths, such as
those inspired by $q-$deformed mechanics \cite{KLIMEK.93}. This does not
alter any of the principles we employ and simply changes the details of the
system function used and hence the sort of invariants of the motion we can
find. In this paper we are interested in treating time homogeneously, and so
we choose a temporal lattice with a constant fundamental time interval $T$.
Our proposed solution for the virtual path in point particle mechanics is to
take the geodesic or shortest geometric path from $\mathbf{q}_n$ to $\mathbf{%
q}_{n+1}$, the metric being normally the Euclidean one in physical space
(not in co-ordinate space). This prescription will normally provides us with
a unique system function from a given Lagrangian. Moreover, it should be
applicable to configuration spaces with curvature and is a co-ordinate frame
independent concept. It allows us to construct a gauge-invariant discrete
time prescription for electrodynamics, with a suitable modification. In 
\textit{Paper II} of this series we shall show that we can apply this
prescription successfully to field theories also. There may be important
cases where the chosen virtual path is not a linear interpolation. This
occurs for charged fields in the next paper in the series. In such cases,
additional requirements such as gauge invariance will influence the choice
of virtual path.

To illustrate the procedure, consider a non-relativistic particle with
position vector $\mathbf{x}$ and Lagrangian $L(\mathbf{x},\mathbf{\dot{x}}%
,t) $. Then the virtual path $\mathbf{\tilde{x}}_n$ taken between $\mathbf{x}%
_n$ and $\mathbf{x}_{n+1}$ is given by 
\begin{equation}
\mathbf{\tilde{x}}_n=\lambda \mathbf{x}_{n+1}+\bar{\lambda}\mathbf{x}_n,
\label{path}
\end{equation}
where $0\leq \lambda \leq 1$ and $\bar{\lambda}\equiv 1-\lambda $. With this
choice of virtual path the time derivative become a difference operator.
Specifically, we find 
\begin{equation}
\mathbf{\tilde{v}}_n\equiv \frac d{d\tilde{t}}\mathbf{\tilde{x}}_n=\frac{%
\mathbf{x}_{n+1}-\mathbf{x}_n}T,  \label{vel}
\end{equation}
where we define 
\begin{equation}
\tilde{t}_n\equiv \lambda t_{n+1}+\bar{\lambda}t_n=t_n+\lambda T.
\label{time}
\end{equation}
Then we construct the system function via the rule 
\begin{equation}
F^n=T\int_0^1d\lambda \,L(\mathbf{\tilde{x}}_n,\mathbf{\tilde{v}}_n,\tilde{t}%
_n).
\end{equation}
The use of this integration does not imply that continuous time is regarded
as meaningful in the context of discrete time mechanics. We are interested
only in the results, not in the means of obtaining these results. A useful
analogy is with the use of classical mechanics to set up quantum mechanical
models. Once we have found our quantum theory, we need no longer to regard
the classical model which generated it as any more than some approximation
useful in some circumstances. Our prescription allows us to embed into our
system function fundamental properties such as gauge invariance and other
symmetries of importance to physics.

If the Lagrangian is a real analytic function of its arguments then we may
make a Taylor expansion about $\mathbf{x}_n$ and integrate term by term.
This will be valid for Lagrangians which are polynomial functions of $%
\mathbf{x}$ and $\mathbf{\dot{x}}$. In such cases the system function $F^n$
would be given by the formal expression 
\begin{eqnarray}
F^n=T\sum_{m=0}^\infty \frac{T^m(D_n)^m}{(m+1)!}L(\mathbf{x}_n,\mathbf{v}%
_n,t_n),
\end{eqnarray}
where $D_n$ is the operator $\mathbf{v}_n\mathbf{\cdot }\frac \partial
{\partial \mathbf{x}_n}+\frac \partial {\partial t_n}$ and $\mathbf{v}_n$
and $\mathbf{x}_n$ are considered independent at this stage.

Some examples will illustrate the process. For a particle in one dimension
with Lagrangian 
\begin{equation}
L=\frac 12m{\dot{x}}^2-\sum_{r=0}^\infty C_rx^r
\end{equation}
where the $C_r$ are constants, the system function is given formally by 
\begin{equation}
F^n=\frac{m(x_{n+1}-x_n)^2}{2T}-T\sum_{r=0}^\infty \frac{%
C_r(x_{n+1}^{r+1}-x_n^{r+1})}{(r+1)(x_{n+1}-x_n)}.
\end{equation}
For instance, the anharmonic oscillator Lagrangian 
\begin{equation}
L=\frac 12m\dot{x}^2-\frac 12m{\omega }^2x^2-\frac 14m\lambda x^4
\end{equation}
gives the system function 
\begin{equation}
F^n=\frac{m(x_{n+1}-x_n)^2}{2T}-\frac{Tm{\omega }^2}6\frac{(x_{n+1}^3-x_n^3)%
}{(x_{n+1}-x_n)}-\frac{Tm\lambda }{20}\frac{(x_{n+1}^5-x_n^5)}{(x_{n+1}-x_n)}%
.
\end{equation}
This differs from the anharmonic oscillator system function discussed in 
\cite{JAROSZKIEWICZ.94A}, which illustrates the general problem with
discrete time mechanics. There may be many possible discretisation of a
given continuous time system, all of which lead back to the continuous time
theory when we take appropriate limits. The principle specified above gives
us a unique discretisation (subject to choice of virtual path).

For the coulombic potential problem in three spatial dimensions, the
Lagrangian 
\begin{equation}
L=\frac m2\mathbf{\dot{x}\cdot \dot{x}}+\frac \gamma {|\mathbf{x}|}
\label{coul}
\end{equation}
with virtual path $\left( \ref{path}\right) $ gives the system function 
\begin{eqnarray}
F^n &=&\frac{m(\mathbf{x}_{n+1}-\mathbf{x}_n)^2}{2T}  \nonumber  \label{dd}
\\
&&+\frac{\gamma T}{|\mathbf{x}_{n+1}-\mathbf{x}_n|}ln\left\{ \frac{\mathbf{x}%
_{n+1}\mathbf{\cdot }(\mathbf{x}_{n+1}-\mathbf{x}_n)+|\mathbf{x}_{n+1}||%
\mathbf{x}_{n+1}-\mathbf{x}_n|}{\mathbf{x}_n\mathbf{\cdot }(\mathbf{x}_{n+1}-%
\mathbf{x}_n)+|\mathbf{x}_n||\mathbf{x}_{n+1}-\mathbf{x}_n|}\right\} .
\end{eqnarray}
This system function leads to Cadzow's equations of motion which preserve
the discrete time analogue of orbital angular momentum. This system function
is markedly different in form to the original coulombic Lagrangian $\left( 
\ref{coul}\right) $ but if we consider trajectories for which we may write $%
\mathbf{x}_n\equiv \mathbf{r}_n,\;\mathbf{x}_{n+1}\equiv $ $\mathbf{r}_n+$ $T%
\mathbf{v}_n\mathbf{+}O\left( T^2\right) $ for each $n$, then 
\begin{equation}
\lim_{T\rightarrow 0}\left\{ \frac{F^n}T\right\} =\frac{_1}{^2}m\mathbf{v}_n%
\mathbf{\cdot v}_n\mathbf{+}\frac \gamma {\left| \mathbf{r}_n\right| },
\end{equation}
which corresponds with $\left( \ref{coul}\right) $. However, it should be
kept in mind that there will be many discrete time trajectories for which
this limit cannot be taken. For example, there may be trajectories where the
particle repeatedly flips between two fixed positions only. This may happen
with the discrete time harmonic oscillator, for example, and no limit such
as the one discussed above exists for such a trajectory. Discrete time
mechanics is inherently richer in its set of possible trajectories than
continuous time mechanics.

In general, Cadzow's equations lead to an implicit equation for $x_{n+1}$
involving $x_n$ and $x_{n-1}$, although for certain systems such as the
harmonic oscillator discussed below we may solve Cadzow's equation to find $%
x_{n+1}$ explicitly. The situation is analogous to what happens in computer
simulations of partial differential equations where not all equations give $%
x_{n+1}$ explicitly. In such cases we must use numerical techniques to solve
for the $x_{n+1}$ in the classical theory. It is a special feature of our
approach that our equations of motion involve only $x_{n-1}$, $x_n$, and $%
x_{n+1}$, which is not always the case with finite difference schemes used
to approximate differential equations.\ 

\section{Invariants of the motion}

It is possible to find a discrete time analogue of Noether's theorem in the
case of continuous symmetries along the lines considered by Maeda \cite
{MAEDA.81}. We shall refer to constants of the motion found by this theorem
as \textit{Maeda-Noether invariants}\emph{.} Consider a system function $%
F^n\equiv F(\mathbf{q}_n,\mathbf{q}_{n+1})$ which is invariant to some point
transformation $\mathbf{q}_n\rightarrow \mathbf{q}_n^{\prime }=\mathbf{q}%
_n+\delta \mathbf{q}_n.$ Then we may write 
\begin{eqnarray}
0=\delta F^n &=&\frac{\partial F^n}{\partial \mathbf{q}_n}\mathbf{\cdot }%
\delta \mathbf{q}_n+\frac{\partial F^n}{\partial \mathbf{q}_{n+1}}\mathbf{%
\cdot }\delta \mathbf{q}_{n+1}  \nonumber \\
\, &&\stackunder{c}{=}\frac{\partial F^n}{\partial \mathbf{q}_n}\mathbf{%
\cdot }\delta \mathbf{q}_n-\frac{\partial F^{n+1}}{\partial \mathbf{q}_{n+1}}%
\mathbf{\cdot }\delta \mathbf{q}_{n+1}
\end{eqnarray}
using Cadzow's equation of motion. From this we deduce that the quantity $%
C^n\equiv \frac{\partial F^n}{\partial \mathbf{q}_n}\mathbf{\cdot }\delta 
\mathbf{q}_n$ will be conserved along dynamical trajectories, that is, 
\begin{equation}
C^n\stackunder{c}{=}C^{n+1}.
\end{equation}

This construction does not allow us to construct an analogue of the
Hamiltonian in the case of conserved systems because in our formulation we
are not allowed to make infinitesimal jumps in time.

Logan \cite{LOGAN.73} gave a method for constructing constants of the motion
which are not necessarily related to symmetries of the system function.
Consider a point transformation 
\begin{equation}
\mathbf{q}_n\rightarrow \mathbf{q}_n^{\prime }=\mathbf{q}_n+\epsilon \mathbf{%
u}_n,  \label{trans}
\end{equation}
where $\epsilon $ is infinitesimal and $\mathbf{u}_n$ is a function of $%
\mathbf{q}_n$ and $\mathbf{q}_{n+1}$. Then 
\begin{equation}
\delta F^n=\epsilon \left\{ \frac{\partial F^n}{\partial \mathbf{q}_n}%
\mathbf{\cdot u}_n+\frac{\partial F^n}{\partial \mathbf{q}_{n+1}}\mathbf{%
\cdot u}\Sb n+1 \\  \endSb \right\} \stackunder{c}{=}\epsilon \frac{\partial
F^n}{\partial \mathbf{q}_n}\mathbf{\cdot u}_n-\epsilon \frac{\partial F^{n+1}%
}{\partial \mathbf{q}_{n+1}}\mathbf{\cdot u}_{n+1},
\end{equation}
on the true trajectories. Now suppose that the transformation $\left( \ref
{trans}\right) $ is such that $\delta F^n$ can be written in the form$%
\;\delta F^n=\epsilon v_{n+1}-\epsilon v_n,$ where $v_n=v(\mathbf{q}_n).$
Then we immediately deduce that the quantity 
\begin{equation}
C^n\equiv \frac{\partial F^n}{\partial \mathbf{q}_n}\mathbf{\cdot u}_n+v_n
\end{equation}
is conserved over the classical trajectories. Such a constant of the motion
will be referred to as a \textit{Logan invariant.}\ 

\section{The discrete time harmonic oscillator}

\subsection{A Logan invariant for the harmonic oscillator}

The discrete time harmonic oscillator in its generic form is given by the
quadratic system function 
\begin{equation}
F^n=\frac{_1}{^2}\alpha \left( x_n^2+x_{n+1}^2\right) -\beta
x_nx_{n+1},\;\;\;\;\;\beta >0,  \label{gen}
\end{equation}
which gives the Cadzow's equation of motion 
\begin{equation}
x_{n+1}\stackunder{c}{=}2\eta x_n-x_{n-1},\;\;\;\;\eta =\frac \alpha \beta .
\label{eqn}
\end{equation}
A Logan invariant of the motion is found to be 
\begin{equation}
C^n\equiv \frac{_1}{^2}\beta \left( x_n^2+x_{n+1}^2\right) -\alpha
x_nx_{n+1}.  \label{Loganho}
\end{equation}

\subsection{Limiting behaviour}

In this subsection we show how to solve the equation of motion $\left( \ref
{eqn}\right) $ and determine the behaviour of the oscillator as the discrete
time tends to infinity. First we define the variables 
\begin{equation}
a_n^{\pm }\equiv x_n-\mu ^{\pm }x_{n+1},
\end{equation}
which will become the analogues of annihilation and creation operators in
quantum theory. The constants $\mu ^{\pm }$ are chosen to satisfy the
condition 
\begin{equation}
a_n^{\pm }\stackunder{c}{=}\mu ^{\pm }a_{n-1}^{\pm }  \label{cond}
\end{equation}
under the equation of motion $\left( \ref{eqn}\right) ,$ which implies 
\begin{equation}
a_n^{\pm }\stackunder{c}{=}\left( \mu ^{\pm }\right) ^na_0^{\pm }.
\end{equation}
Condition $\left( \ref{cond}\right) $ gives 
\begin{equation}
\mu ^{\pm }=\eta \pm \sqrt{\eta ^2-1}.
\end{equation}
We note that $\mu ^{+}\mu ^{-}=1.$ The Logan invariant $\left( \ref{Loganho}%
\right) $ is given by 
\begin{equation}
C^n=\frac{_1}{^2}\beta a_n^{+}a_n^{-},
\end{equation}
which is a constant of the motion by inspection and is a form of great value
in discrete time field theory.

The complete solution to the problem is now readily obtained and given by 
\begin{equation}
x_n\stackunder{c}{=}\frac{\left[ \left( \mu ^{+}\right) ^n-\left( \mu
^{-}\right) ^n\right] x_1+\left[ \left( \mu ^{-}\right) ^{n-1}-\left( \mu
^{+}\right) ^{n-1}\right] x_0}{\left( \mu ^{+}-\mu ^{-}\right) },\;\;\;\eta
^2\neq 1.  \label{sol}
\end{equation}
For the case when $\eta ^2<1$ we write $\eta =\cos \left( \theta \right) $
and then we find 
\begin{equation}
x_n\stackunder{c}{=}\frac{\sin \left( n\theta \right) x_1-\sin \left( \left(
n-1\right) \theta \right) x_0}{\sin \left( \theta \right) },
\end{equation}
whereas for $\eta ^2>1$ we write $\eta =\cosh \left( \chi \right) $
(assuming $\eta >1),$ and then 
\begin{equation}
\;x_n\stackunder{c}{=}\frac{\sinh \left( n\chi \right) x_1-\sinh \left(
\left( n-1\right) \chi \right) x_0}{\sinh \left( \chi \right) }.
\end{equation}
and similarly for $\eta <-1.$ The crucial result is that bounded, elliptic
behaviour occurs when $\eta ^2<1$ whereas unbounded (hyperbolic) behaviour
occurs when $\eta ^2>1.$ This result gives a natural cutoff to particle
momentum in field theory, as shown in \textit{Paper II}.

The readily solved case when $\eta ^2=1$ corresponds to the free particle
and will be referred to as the \textit{parabolic }case. When $\eta ^2>1$ the
system will be said to be \textit{hyperbolic} and \textit{elliptic} when $%
\eta ^2<1$.

For the case $\eta ^2<1$ it is useful to define $\mu =\eta +i\sqrt{1-\eta ^2%
\text{ }}$ and 
\begin{equation}
a_n\equiv \mu ^n\left[ x_{n+1}-\mu x_n\right] ,\;\;\;a_n^{*}\equiv \mu
^{-n}\left[ x_{n+1}-\mu ^{-1}x_n\right] ,  \label{ladder}
\end{equation}
the advantage being that these are constants of the motion, viz 
\begin{equation}
a_n\stackunder{c}{=}a_{n-1},\;\;\;a_n^{*}\stackunder{c}{=}a_{n-1}^{*}.
\end{equation}
These are useful in the construction of particle states in quantum field
theory because they correspond to annihilation and creation operators in the
Schr\"{o}dinger picture.

\subsection{The Newtonian oscillator}

Using the methods outlined in \S $3$ the continuous time Lagrangian for the
Newtonian harmonic oscillator 
\begin{equation}
L=\frac{{}_1}{{}^2}m{\dot{x}}^2-\frac{{}_1}{{}^2}m\omega ^2x^2  \label{NHO}
\end{equation}
gives the system function 
\begin{equation}
F^n=\frac{m(x_{n+1}-x_n)^2}{2T}-\frac{Tm\omega ^2}%
6(x_{n+1}^2+x_{n+1}x_n+x_n^2),  \label{ho}
\end{equation}
which is equivalent to $\left( \ref{gen}\right) .$ The equation of motion is
given by 
\begin{equation}
\frac{\left( x_{n+1}-2x_n+x_{n-1}\right) }{T^2}\stackunder{c}{=}-\omega ^2%
\frac{\left( x_{n+1}+4x_n+x_{n-1}\right) }6,  \label{hoeqm}
\end{equation}
which is equivalent to $\left( \ref{eqn}\right) $ with the identification $%
T^2\omega ^2=6(1-\eta )/(2+\eta )$, which means 
\begin{equation}
\eta =\frac{6-2T^2\omega ^2}{6+T^2\omega ^2}.
\end{equation}
Using the results of the previous subsection, we deduce that elliptic
behaviour occurs only when the time $T$ satisfies the condition 
\begin{equation}
0<T\omega <2\sqrt{3}.
\end{equation}
An equivalent result is found in particle field theory, giving a natural
cutoff for particle momentum.\ 

\subsection{Harmonic recurrence}

We may understand the relationship between Hamilton's principal function for
the interval $\left[ 0,T\right] $ and the system function by explicitly
evaluating the former for the continuous time harmonic oscillator Lagrangian 
$\left( \ref{NHO}\right) $. We find 
\begin{equation}
S^n\left( T\right) =\frac{m\omega }{2\sin (\omega T)}\left[
(x_{n+1}^2+x_n^2)\cos (\omega T)-2x_nx_{n+1}\right] ,  \label{ham}
\end{equation}
and comparing this with $\left( \ref{ho}\right) $ we find 
\begin{equation}
S^n\left( T\right) =F^n+O\left( T^3\right) .
\end{equation}
We expect a similar relation to exist in the general case, but different
potentials will modify the precise details.

There is an apparent problem with $\left( \ref{ham}\right) $ whenever the
time interval $T$ satisfies the condition $\omega T=r\pi $ , $r=1,2,...$
because the denominator $\sin \left( \omega T\right) $ vanishes at such
times. This problem is an artefact of our representation of $S^n$, because
the definition of the principal function as a line integral over a finite
contour of a bounded integrand means that $S^n$ cannot diverge. The
resolution of this apparent paradox is that at the \textit{recurrence times} 
$T=r\pi /\omega $ the endpoints $x_n$ and $x_{n+1}$ are no longer
independent but are related by 
\begin{equation}
x_{n+1}=\left( -1\right) ^rx_n.  \label{recur}
\end{equation}

The physical interpretation of recurrence is simple. The harmonic oscillator
has a fundamental period $P=2\pi /\omega $ , independent of the initial
conditions.

An important construction for the harmonic oscillator are the variables $A_n$%
, $A_n^{*}$ defined by 
\begin{eqnarray}
A_n &\equiv &\frac{ie^{in\theta }}{\sin \left( \theta \right) }\left[
x_{n+1}-e^{i\theta }x_n\right] ,  \nonumber \\
A_n^{*} &\equiv &\frac{-ie^{-in\theta }}{\sin \left( \theta \right) }\left[
x_{n+1}-e^{-i\theta }x_n\right] ,\;\;\;\theta \equiv \omega T.
\label{laddera}
\end{eqnarray}
These are constants of the motion, i.e. $A_n\stackunder{c}{=}A_{n+1}$ and
are independent of $T$. This means that recurrence must occur so as to
cancel the zero of the denominator in $\left( \ref{ham}\right) $ at the
recurrence times. To see what happens explicitly, we may invert the
equations $\left( \ref{laddera}\right) $ to find 
\begin{equation}
x_n=\frac{{}_1}{{}^2}\left[ e^{in\theta }A_n^{*}+e^{-in\theta }A_n\right]
,\;\;\;x_{n+1}=\frac{{}_1}{{}^2}\left[ e^{i(n+1)\theta
}A_n^{*}+e^{-i(n+1)\theta }A_n\right] ,
\end{equation}
so that at the recurrence times $T=r\pi /\omega $ we have 
\begin{equation}
x_n=\frac{\left( -1\right) ^{nr}}2\left[ A_n^{*}+A_n\right] ,
\end{equation}
from which we deduce $\left( \ref{recur}\right) .$

In terms of $A_n$ and $A_n^{*}$ the principal function can be written as 
\begin{equation}
S^n\left( T\right) =-\frac{m\omega \sin \left( \theta \right) }4\left[
e^{i\left( 2n+1\right) \theta }A_n^{*2}+e^{-i\left( 2n+1\right) \theta
}A_n^2\right] ,
\end{equation}
a form which shows clearly that the principal function is not singular.
Moreover, we see that at the recurrence times 
\begin{equation}
S^n\left( \frac{r\pi }\omega \right) =0,\hspace{0.5in}r=1,2,...
\end{equation}

It is a significant feature of the discrete time harmonic oscillator that it
does not involve recurrence phenomena in this particular way, as no apparent
singularities occur in the system function $\left( \ref{ho}\right) $. This
emphasises that discrete time mechanics is not equivalent to continuous time
mechanics.\ 

\section{Electrodynamics: test particles}

$~~$We now consider the case of electrically charged particles interacting
with electromagnetic fields. A more complete discussion of discrete time
Maxwell's equations is given in Paper $II$ of this series. Here we discuss
only the case of test particles which are affected by external electric and
magnetic fields but do not affect them. We shall find the Cadzow equation of
motion for such particles and show that in our prescription electric charge
is conserved.

Consider a non-relativistic charged test particle of mass $m$ in external
electromagnetic potentials. The continuous time Lagrangian for such a system
is given by 
\begin{equation}
L_{EM}=\frac{_1}{^2}m\mathbf{\dot{x}\cdot \dot{x}}+q\mathbf{\dot{x}\cdot A}(%
\mathbf{x},t)-q\phi (\mathbf{x},t),  \label{Lem}
\end{equation}
where $q$ is the charge of the particle. This Lagrangian is not gauge
invariant but the equations of motion are gauge invariant, because under the
gauge transformation 
\begin{eqnarray}
\phi \rightarrow \phi ^{\prime } &\equiv &\phi +{\partial }_t\chi , 
\nonumber \\
\mathbf{A}\rightarrow \mathbf{A^{\prime }} &\equiv &\mathbf{A}-\nabla \chi ,
\end{eqnarray}
the action integral transforms according to the rule 
\begin{equation}
A_{if}\rightarrow A_{if}^{\prime }\equiv A_{if}-\left[ q\chi \right]
_{t_i}^{t_f},  \label{linesum}
\end{equation}
that is, the change in the action integral occurs only at the end points. If
this property is preserved in any discretised version of electrodynamics
then the equations of motion should be gauge invariant. Our prescription for
calculating the system function from the Lagrangian does indeed preserve
this property and therefore our discrete time equations of motion are gauge
invariant.

The first step is to construct discrete time electromagnetic potentials.
These are discussed in full detail in Paper $II$, but the basic properties
are the following. The magnetic vector potential $\mathbf{A}$ differs to the
scalar potential $\phi $ in that the former is defined at temporal lattice
sites whereas the latter is defined on the links between these sites. If $%
\mathbf{A}_n\left( \mathbf{x}\right) $ is the value of the vector potential
at time $n$ at position $\mathbf{x}$, and $\phi _n\left( \mathbf{x}\right) $
is the scalar potential on the link at position $\mathbf{x}$ from time $n$
to time $n+1,$ then under a discrete time gauge transformation we have 
\begin{eqnarray}
\phi _n^{\prime }\left( \mathbf{x}\right) &=&\phi _n\left( \mathbf{x}\right)
+\frac{\chi _{n+1}\left( \mathbf{x}\right) -\chi _n\left( \mathbf{x}\right) }%
T,  \nonumber \\
\mathbf{A}_n^{\prime }\left( \mathbf{x}\right) &=&\mathbf{A}_n\left( \mathbf{%
x}\right) -\nabla \chi _n\left( \mathbf{x}\right) ,
\end{eqnarray}
where $\chi _n\left( \mathbf{x}\right) $ is the value of the gauge
transformation function at time $n$ and position $\mathbf{x}$. The electric
and magnetic fields are defined by 
\begin{eqnarray}
\mathbf{E}_n\left( \mathbf{x}\right) &=&-\nabla \phi _n\left( \mathbf{x}%
\right) -\frac{\mathbf{A}_{n+1}\left( \mathbf{x}\right) -\mathbf{A}_n\left( 
\mathbf{x}\right) }T,  \nonumber \\
\mathbf{B}_n\left( \mathbf{x}\right) &=&\nabla \times \mathbf{A}_n\left( 
\mathbf{x}\right) .
\end{eqnarray}
These are discrete time gauge invariant. By inspection the electric field is
associated with temporal links whereas the magnetic field is associated with
temporal sites.

In order to apply our discretisation prescription to (\ref{Lem}) we specify
the virtual paths between times $t_n$ and $t_{n+1}$ to be given by 
\begin{eqnarray}
\mathbf{\tilde{x}}_n &\equiv &\lambda \mathbf{x}_{n+1}+\bar{\lambda}\mathbf{x%
}_n,  \nonumber \\
\mathbf{\tilde{A}}_n\left( \mathbf{\tilde{x}}_n\right) &\equiv &\lambda 
\mathbf{A}_{n+1}\left( \mathbf{\tilde{x}}_n\right) +\bar{\lambda}\mathbf{A}%
_n\left( \mathbf{\tilde{x}}_n\right) ,  \nonumber \\
\tilde{\phi}_n\left( \mathbf{\tilde{x}}_n\right) &\equiv &\phi _n\left( 
\mathbf{\tilde{x}}_n\right)
\end{eqnarray}
and then the system function is given by 
\begin{equation}
F^n=\frac{m(\mathbf{x}_{n+1}-\mathbf{x}_n)^2}{2T}+q(\mathbf{x}_{n+1}-\mathbf{%
x}_n)\mathbf{\cdot }\int_0^1d\lambda \,\mathbf{\tilde{A}}_n\left( \mathbf{%
\tilde{x}}_n\right) -Tq\int_0^1d\lambda \,\tilde{\phi}_n\left( \mathbf{%
\tilde{x}}_n\right) .  \label{Sysfun}
\end{equation}

Under a gauge transformation we find 
\begin{equation}
F^{n\prime }=F^n+q\chi _n\left( \mathbf{x}_n\right) -q\chi _{n+1}\left( 
\mathbf{x}_{n+1}\right)
\end{equation}
and so the action sum $A^N\equiv \sum\limits_{n=0}^{N-1}F^n\;$changes
according to the rule 
\begin{equation}
A^{N\prime }=A^N+q\chi _0-q\chi _N,
\end{equation}
in agreement with $\left( \ref{linesum}\right) .\;$ Therefore we expect
Cadzow's equations of motion obtained from $\left( \ref{Sysfun}\right) $ to
be gauge invariant.

In general, the integrals over the external electromagnetic potentials in (%
\ref{Sysfun}) give complicated equations of motion and we will normally have
only an implicit equation for $\mathbf{x}_{n+1}$, which however will be
gauge invariant. We find 
\begin{eqnarray}
&&\frac{m\left( \mathbf{x}_{n+1}-2\mathbf{x}_n+\mathbf{x}_{n-1}\right) }{T^2}%
\stackunder{c}{=}q\int_0^1d\lambda \,\left\{ \bar{\lambda}\mathbf{E}_n\left( 
\mathbf{\tilde{x}}_n\right) +\lambda \mathbf{E}_{n-1}\left( \mathbf{\tilde{x}%
}_{n-1}\right) \right\}  \nonumber \\
&&\;\;\;\;\;\;\;\;\;\;\;\;\;\;\;\;\;\;\;\;\;\;\;\;\;\;+q\frac{\left( \mathbf{%
x}_{n+1}-\mathbf{x}_n\right) }T\times \int_0^1d\lambda \,\left\{ \lambda 
\bar{\lambda}\mathbf{B}_{n+1}\left( \mathbf{\tilde{x}}_n\right) +\bar{\lambda%
}^2\mathbf{B}_n\left( \mathbf{\tilde{x}}_n\right) \right\}  \nonumber \\
&&\;\;\;\;\;\;\;\;\;\;\;\;\;\;\;\;\;\;\;\;\;\;\;\;\;\;+q\frac{\left( \mathbf{%
x}_n-\mathbf{x}_{n-1}\right) }T\times \int_0^1d\lambda \,\left\{ \lambda ^2%
\mathbf{B}_n\left( \mathbf{\tilde{x}}_{n-1}\right) +\bar{\lambda}\lambda 
\mathbf{B}_{n-1}\left( \mathbf{\tilde{x}}_{n-1}\right) \right\} .  \nonumber
\\
&&
\end{eqnarray}
In the limit $T\rightarrow 0$ we recover the usual Lorentz force law 
\begin{equation}
m\mathbf{\ddot{x}}\stackunder{c}{=}q\mathbf{E+}q\mathbf{\dot{x}\times B}
\end{equation}
for those trajectories where the limit exits.

The charge density $\rho _n\left( \mathbf{x}\right) $ and current density $%
\mathbf{j}_n\left( \mathbf{x}\right) $ are defined by the following
functional derivatives with respect to the electromagnetic potentials of the
action sum $A^N:$%
\begin{equation}
\rho _n\left( \mathbf{x}\right) \equiv \frac{-1}T\frac \delta {\delta \phi
_n\left( \mathbf{x}\right) }A^N,\;\;\;\;\;\mathbf{j}_n\left( \mathbf{x}%
\right) \equiv \frac 1T\frac \delta {\delta \mathbf{A}_n\left( \mathbf{x}%
\right) }A^N.
\end{equation}
We find 
\begin{eqnarray}
\rho _n\left( \mathbf{x}\right) &=&q\int_0^1d\lambda \,\delta ^3\left( 
\mathbf{\tilde{x}}_n-\mathbf{x}\right) ,  \nonumber \\
\mathbf{j}_n\left( \mathbf{x}\right) &=&q\frac{\left( \mathbf{x}_{n+1}-%
\mathbf{x}_n\right) }T\int_0^1d\lambda \,\bar{\lambda}\,\delta ^3\left( 
\mathbf{\tilde{x}}_n-\mathbf{x}\right)  \nonumber \\
&&+q\frac{\left( \mathbf{x}_n-\mathbf{x}_{n-1}\right) }T\int_0^1d\lambda
\,\lambda \,\delta ^3\left( \mathbf{\tilde{x}}_{n-1}-\mathbf{x}\right) ,
\end{eqnarray}
which satisfy the discrete time analogue of the equation of continuity 
\[
\frac{\rho _n\left( \mathbf{x}\right) -\rho _{n-1}\left( \mathbf{x}\right) }%
T+\nabla \mathbf{\cdot j}_n\left( \mathbf{x}\right) =0. 
\]
\ 

\section{The Grassmannian oscillator}

We may apply our methods to the Grassmannian oscillator system, which serves
as a prototype model for the Dirac equation studied in \textit{Paper II}.
Our model consists of one complex anticommuting degree of freedom $\psi $
with equation of motion 
\begin{equation}
i\dot{\psi}\stackunder{c}{=}\omega \psi ,\;\;\;i\dot{\psi}^{*}\stackunder{c}{%
=}-\omega \psi ^{*}.  \label{fermi}
\end{equation}
The Lagrangian giving these equations is 
\begin{equation}
L=\frac{_1}{^2}i\psi ^{*}\dot{\psi}-\frac{_1}{^2}i\dot{\psi}^{*}\psi -\omega
\psi ^{*}\psi .  \label{fermil}
\end{equation}
The equations $\left( \ref{fermi}\right) \;$imply the harmonic oscillator
equations of motion 
\begin{equation}
\frac{d^2}{dt^2}\psi \stackunder{c}{=}-\omega ^2\psi ,\;\;\;\frac{d^2}{dt^2}%
\psi ^{*}\stackunder{c}{=}-\omega ^2\psi ^{*}.  \label{hoq}
\end{equation}

Now consider discretisation using the linear virtual paths 
\begin{equation}
\tilde{\psi}=\lambda \psi _{n+1}+\bar{\lambda}\psi _n,\;\;\;\tilde{\psi}%
^{*}=\lambda \psi _{n+1}^{*}+\bar{\lambda}\psi _n^{*}.  \label{vpaths}
\end{equation}
Using $\left( \ref{fermil}\right) \;$we find the system function is 
\begin{eqnarray}
F^n &=&\frac{_1}{^2}i\left[ \psi _n^{*}\psi _{n+1}-\psi _{n+1}^{*}\psi
_n\right]  \nonumber \\
&&-\omega T\frac{\left\{ 2\psi _{n+1}^{*}\psi _{n+1}+\psi _n^{*}\psi
_{n+1}+\psi _{n+1}^{*}\psi _n+2\psi _n^{*}\psi _n\right\} }6,
\end{eqnarray}
which leads to the equation of motion 
\begin{equation}
i\frac{\left( \psi _{n+1}-\psi _{n-1}\right) }{2T}\stackunder{c}{=}\omega 
\frac{\left( \psi _{n+1}+4\psi _n+\psi _{n-1}\right) }6.  \label{f2}
\end{equation}
and similarly for the complex conjugate. It seems not possible to use ($\ref
{f2})$ to obtain the discretisation $\left( \ref{hoeqm}\right) $ of the
harmonic oscillator unless we change the virtual paths or renormalise the
frequency $\omega $. However, we can readily show that $\left( \ref{f2}%
\right) $ implies harmonic oscillator behaviour by the following method.

First, rewrite $\left( \ref{f2}\right) $ as 
\begin{equation}
\left( -3i+\theta \right) \psi _{n+1}+\left( 3i+\theta \right) \psi _{n-1}%
\stackunder{c}{=}-4\theta \psi _n,
\end{equation}
where $\theta \equiv \omega T.$ If we define 
\begin{equation}
\nu \equiv \frac{3i+\theta }{\sqrt{9+\theta ^2}},\;\;\;\nu ^{-1}\equiv \frac{%
-3i+\theta }{\sqrt{9+\theta ^2}}
\end{equation}
and shift the degrees of freedom according to the rule 
\begin{equation}
\psi _n\equiv \nu ^n\phi _n
\end{equation}
then the variables $\phi _n$ satisfy the discrete time oscillator equation 
\begin{equation}
\phi _{n+1}+\phi _{n-1}\stackunder{c}{=}2\eta \phi _n,
\end{equation}
where $\eta =\frac{-2\theta }{\sqrt{9+\theta ^2}}.$ We find 
\begin{equation}
\mu \equiv \eta +i\sqrt{1-\eta ^2}=\frac{-2\theta +i\sqrt{9-3\theta ^2}}{%
\sqrt{9+\theta ^2}},
\end{equation}
from which we deduce the upper limit $\omega T<\sqrt{3}$ for elliptic
behaviour in the system. This is exactly one half of the upper limit found
for the bosonic discrete time oscillator.\ 

\section{First quantisation}

We now discuss the possibility of quantising our classical discrete time
mechanics. If we denote the process of quantisation by the symbol $\mathcal{Q%
}$ and the process of discretisation by the symbol $\mathcal{D}\;$then the
question arises, do these processes commute, i.e. does $\mathcal{QD}%
\stackrel{?}{=}\mathcal{DQ}$. In other words, does it matter if we
discretise the quantum theory of some system with classical Lagrangian
rather than quantise the discretised version of the same classical
Lagrangian?

There is \textit{a priori} no reason to expect these processes to commute.
For one thing, we have not yet decided what $\mathcal{Q}$ might mean. Also,
there are some aspects of $\mathcal{D}$ such as the lack of a Hamiltonian
which makes a dynamical quantum theory of discrete time point particle
mechanics problematical. Fortunately, there are some concepts from the
conventional $\mathcal{Q}~$programme which are useful and appear to survive $%
\mathcal{D}$. We shall comment on some of these aspects now and then
consider the harmonic oscillator quantisation process $\mathcal{QD}$ in some
detail in the following sections, reserving a discussion of the $\mathcal{DQ}
$ process for Paper $II$.

In the following we discuss a system consisting of a point particle in one
dimension. Generalisation to more degrees of freedom is straightforward and
we shall use the Dirac bra-ket notation for convenience.

\subsection{Basics}

In our approach to quantisation, we shall follow all the standard principles
of orthodox quantum mechanics in the main. This means we face the same
issues of rigour and interpretation as orthodox quantum theory. We shall not
comment on those in general. We discuss below those aspects where
discretisation of time requires some additional emphasis or comment.

\textbf{Proposition 1: }\textit{Physical states of a quantum system
correspond one-to-one to rays in a separable Hilbert space} $\mathcal{H}$. 
\textit{A physical state vector} $|\phi \rangle $ \textit{will be in general
normalised to unity, viz} 
\begin{equation}
\langle \phi |\phi \rangle =1.
\end{equation}

\textbf{Proposition 2: }\textit{For each integer time} $n$ \textit{(or more
accurately, at each co-ordinate time} $nT$\textit{),} $\mathcal{H}$ \textit{%
is spanned by an improper basis} $\mathcal{B}^n\equiv \left\{ |x,n\rangle
:x\in \Re \right\} $\textit{, the elements of which satisfy the relation} 
\begin{equation}
\langle x,n|y,n\rangle =\delta \left( x-y\right) .
\end{equation}

The resolution of the identity operator in $\mathcal{H}$ is 
\begin{equation}
\hat{I}_{\mathcal{H}}=\int dx\,|x,n\rangle \langle x,n|,  \label{res}
\end{equation}
which holds for each $n$.

Given a physical state $|\psi \rangle $ in $\mathcal{H}$ we may write for
each $n$ 
\begin{equation}
|\psi \rangle =\int dx\,\psi _n\left( x\right) |x,n\rangle ,
\end{equation}
where $\psi _n\left( x\right) $ is the wavefunction at time $n$, with the
property that 
\begin{equation}
\int dx\,|\psi _n\left( x\right) |^2=1,
\end{equation}
assuming normalisation to unity.

\textbf{Remark 1:} \textit{The Heisenberg picture is being used in the
above. The time dependence of the basis sets} $\mathcal{B}^n$ \textit{allows
us to use the Schr\"{o}dinger picture, discussed below.}

\textbf{Remark 2: }\textit{All the usual principles of quantum mechanics
concerning the interpretation of the states in the Hilbert space apply here.
For example, a) the superposition principle and its interpretation according
to standard quantum mechanics holds and b) given two physical states} $|\phi
\rangle $, \TEXTsymbol{\vert}$\psi \rangle $ \textit{then the inner product} 
$\langle \phi |\psi \rangle $ \textit{gives the conditional transition
amplitude for the system to be found in state} \TEXTsymbol{\vert}$\phi
\rangle $\textit{, given it is in state} \TEXTsymbol{\vert}$\psi \rangle $.

\textbf{Definition 1: }\textit{An operator} $\hat{A}$ \textit{diagonal with
respect to} $\mathcal{B}^n$ \textit{is one which can be written in the form} 
\begin{equation}
\hat{A}=\int dx\,|x,n\rangle A(x,\partial _x)\langle x,n|,
\end{equation}
\textit{where the component operator} $A(x,\partial _x)$ \textit{is some
differential operator of finite order}.

The action of such an operator on a typical state $|\psi \rangle $ is given
by 
\begin{equation}
|A\psi \rangle \equiv \hat{A}|\psi \rangle =\int dx\,|x,n\rangle
A(x,\partial _x)\psi _n\left( x\right) 
\end{equation}
with matrix elements given by 
\begin{equation}
\langle \phi |A\psi \rangle =\int dx\,\phi ^{*}\left( x\right) A(x,\partial
_x)\psi _n\left( x\right) .
\end{equation}

\textbf{Remark 3: }\textit{The wavefunctions of the theory are elements of} $%
\mathcal{L}^2\left( \Re \right) $\textit{, the space of square integrable
functions on} $\Re ,$\textit{\ and the operators (including the observables
of the theory) which act on them are usually built up of functions of} $x$ 
\textit{and} $\partial _x.$ \textit{If at a given time} $n$ \textit{the
component operator of some observable diagonal with respect to} $\mathcal{B}%
^{n\text{ }}$\textit{happens to be represented by say a multiple of} $%
\partial _x$ \textit{this carries no implication that the observable is
related to a velocity (which would normally be implied in conventional wave
mechanics, where the momentum operator is represented by} $-i\hbar \partial
_x).$\textit{\ There is no concept of velocity in the normal sense in
discrete time mechanics.}

Keeping in mind the caveats discussed by \cite{STREATER & WIGHTMAN}
concerning hermitian and adjoint operators, we define the hermitian
conjugate or adjoint operator $\hat{A}^{+}$ to have the property that 
\begin{equation}
\langle \phi |A\psi \rangle =\langle \psi |\hat{A}^{+}\phi \rangle ^{*}
\end{equation}
for a dense set of physical states. If $\hat{A}$ is diagonal with respect to 
$\mathcal{B}^{n\text{ }}$then assuming we may represent $\hat{A}^{+}$ in the
form 
\begin{equation}
\hat{A}^{+}=\int dx\,|x,n\rangle \tilde{A}(x,\partial _x)\langle x,n|
\end{equation}
for some operator component $\tilde{A}(x,\partial _x)$ we find 
\begin{equation}
\int dx\,\phi _n^{*}\left( x\right) A\left( x,\partial _x\right) \psi
_n\left( x\right) =\int dx\,\left[ \tilde{A}^{*}\left( x,\partial _x\right)
\phi _n^{*}\left( x\right) \right] \psi _n\left( x\right) .
\end{equation}
Assuming that we are permitted to integrate by parts, which will be the case
for normalisable wavefunctions falling off at spatial infinity, we can
readily understand the relationship between $A\left( x,\partial _x\right) $
and $\tilde{A}(x,\partial _x)$. Given the former we can always work out the
latter by integration by parts and vice-versa.

If $A\left( x,\partial _x\right) =\tilde{A}(x,\partial _x)$ then $\hat{A}=%
\hat{A}^{+}$ and the operator is self-adjoint. Physical observables of the
theory which are diagonal with respect to $\mathcal{B}^n$ will be assumed to
have this property.

\subsection{Dynamics}

The dynamical content of the theory is expressed in terms of unitary \textit{%
timestep operators} $\hat{U}_n$, one for each $n$.

\textbf{Proposition 3: }\textit{For each} $n$ \textit{there is a unitary
operator} $\hat{U}_n$ \textit{such that} 
\begin{equation}
|x,n+1\rangle =\hat{U}_n^{\dagger }|x,n\rangle .
\end{equation}

From this we deduce the relations 
\begin{eqnarray}
\langle x,n+1| &=&\langle x,n|\hat{U}_n,  \nonumber \\
|x,n\rangle  &=&\hat{U}_n|x,n+1\rangle , \\
\;\langle x,n| &=&\langle x,n+1|\hat{U}_n^{\dagger }.  \nonumber
\end{eqnarray}

\textbf{Remark 4: }\textit{The operator} $\hat{U}_n$ \textit{provides an
isometry between} $\mathcal{B}^n$ \textit{to} $\mathcal{B}^{n+1}$\textit{,
that is} 
\begin{equation}
\langle x,n+1|y,n+1\rangle =\langle x,n|y,n\rangle =\delta \left( x-y\right)
.
\end{equation}

Using the resolution of the identity (\ref{res}) we may represent the
timestep operators in the non-diagonal form 
\begin{equation}
\hat{U}_n=\int dx|x,n\rangle \langle x,n+1|,\;\;\;\;\;\hat{U}_n^{\dagger
}=\int dx|x,n+1\rangle \langle x,n|.
\end{equation}

We are now in a position to define the fundamental functions of the quantum
theory, the \textit{system amplitudes }$U_n\left( x,y\right) $, defined by 
\begin{eqnarray}
U_n\left( x,y\right)  &\equiv &\langle x,n+1|y,n\rangle =\langle x,n|\hat{U}%
_n|y,n\rangle ,  \nonumber \\
U_n^{*}\left( x,y\right)  &\equiv &\langle y,n|x,n+1\rangle =\langle y,n|%
\hat{U}_n^{\dagger }|x,n\rangle ,
\end{eqnarray}
from which we arrive at the non-diagonal expressions 
\begin{eqnarray}
\hat{U}_n &=&\int dxdy\,|x,n\rangle U_n\left( x,y\right) \langle y,n|, 
\nonumber \\
\hat{U}_n^{\dagger } &=&\int dxdy\,|x,n\rangle U_n^{*}\left( y,x\right)
\langle y,n|.
\end{eqnarray}

\textbf{Remark 5:} \textit{The system amplitudes will not be differential
operators of finite order in general but well-behaved complex valued
functions of two real variables. Neither will they be singular
distributions. They are similar in function to integrated Feynman transition
kernels encountered in the path integral formulation of standard quantum
mechanics.}

The condition that the timestep operators are unitary, viz 
\begin{equation}
\hat{U}_n\hat{U}_n^{\dagger }=\hat{I}_{\mathcal{H}}
\end{equation}
leads to the closure condition 
\begin{equation}
\int dy\,U_n\left( x,y\right) U_n^{*}\left( z,y\right) =\delta \left(
x-z\right)   \label{unit}
\end{equation}
on the system amplitudes.

\textbf{Definition 2: }\textit{A system for which the system amplitudes are
independent of time, such that we may write} 
\begin{equation}
U_n\left( x,y\right) =U\left( x,y\right) 
\end{equation}
\textit{for some} $U\left( x,y\right) $ \textit{and for all} $n$\textit{,
will be said to be} \textit{autonomous}.

\textbf{Definition 3: }\textit{An autonomous system for which the system
amplitude} $U\left( x,y\right) $ \textit{carries the symmetry} 
\begin{equation}
U\left( x,y\right) =U\left( y,x\right) 
\end{equation}
\textit{will be said to be time-reversal invariant}.

\textbf{Remark 6:} \textit{Most of the system amplitudes of interest to us
will be autonomous and time-reversal invariant. This will be so whenever we
construct system amplitudes from system functions which have been obtained
from conventional time-translation invariant and time-reversal invariant
Lagrangians using the virtual path approach discussed above}.

\subsection{The Schr\"{o}dinger picture}

The Heisenberg picture description of the physical states used so far means
that we may write 
\begin{equation}
|\psi \rangle =\int dx\,\psi _{n+1}\left( x\right) |x,n+1\rangle =\int
dx\,\psi _n\left( x\right) |x,n\rangle ,
\end{equation}
from which we deduce 
\begin{equation}
\psi _{n+1}\left( x\right) =\int dy\,U_n\left( x,y\right) \psi _n\left(
y\right) .  \label{aaa}
\end{equation}
From this we see that the system amplitudes play a role analogous to finite
time scattering kernels in conventional quantum mechanics. Equation $\left( 
\ref{aaa}\right) $ is about the closest we come in this theory to something
analogous to a time dependent Schr\"{o}dinger wave equation.

We may set up a formal description in the Schr\"{o}dinger picture as
follows. Given a Heisenberg picture state $|\psi \rangle $ and a knowledge
of the component functions $\psi _n\left( x\right) $, define the sequence of
states 
\begin{equation}
|\psi _{n,m}\rangle \equiv \int dx\,\psi _n\left( x\right) |x,m\rangle ,
\end{equation}
for some chosen time $m$. Then if $|\psi _{m,m}\rangle \equiv |\psi \rangle $
we find 
\begin{equation}
\hat{U}_m|\psi _{m,m}\rangle =|\psi _{m+1,m}\rangle .
\end{equation}
It is straightforward to extend this to jumps over more than one time
interval. This establishes the Schr\"{o}dinger picture in this theory.

\subsection{Position eigenstates}

Given an improper basis $\mathcal{B}^n\equiv \left\{ |x,n\rangle :x\in \Re
\right\} $ we may construct a self-adjoint position operator 
\begin{equation}
\hat{x}_n\equiv \int dx\,|x,n\rangle x\langle x,n|,
\end{equation}
which has the property $_{}$%
\begin{equation}
\hat{x}_n|x,n\rangle =x|x,n\rangle .
\end{equation}

\textbf{Remark 7: }\textit{For convenience we shall follow the traditional
abuse of notation and use the symbol} $x$ \textit{both for the position
operator and for a particular eigenvalue of that operator. It will be clear
normally from the context what is meant whenever a clash of notation occurs.}

The position operators have the merit of being diagonal with respect to the
appropriate basis, that is, $\hat{x}_n$ is diagonal with respect to $%
\mathcal{B}^n.$ The position operators are not necessarily diagonal with
respect to bases at other values of $n$. From the condition 
\begin{equation}
\hat{x}_{n+1}=\hat{U}_n^{\dagger }\hat{x}_n\hat{U}_n,  \label{pos}
\end{equation}
we find 
\begin{equation}
\hat{x}_{n+1}=\int dxdydz\,\,|x,n\rangle U_n^{*}\left( y,x\right)
\,y\,U_n\left( y,z\right) \langle z,n|,
\end{equation}
which is self-adjoint but not necessarily diagonal with respect to $\mathcal{%
B}^n.$

\textbf{Remark 8:} \textit{It is possible for} $\hat{x}_{n+1}$ \textit{to
reduce to diagonal form with respect to the} $\mathcal{B}^n$ \textit{basis
but this depends on the details of the system amplitudes}.

From the above we arrive at the fundamental expression for the commutators: 
\begin{equation}
\left[ \hat{x}_{n+1},\hat{x}_n\right] =\int dxdydz\,\,|x,n\rangle
U_n^{*}\left( y,x\right) \,y\left( z-x\right) U_n\left( y,z\right) \langle
z,n|.  \label{comm}
\end{equation}

\subsection{Normal co-ordinate systems}

In this subsection we discuss the quantisation of a large family of systems
for which the following property holds:

\textbf{Definition 4: }\textit{Co-ordinates for a system for which the right
hand side of the commutator} (\ref{comm}) \textit{is a multiple of the
identity operator for each value of} $n$ \textit{will be called normal}.

\textbf{Remark 9: }\textit{We shall see below that the co-ordinates for the
important system equivalent to the harmonic oscillator are normal.}

A class of system amplitudes for which the co-ordinates are normal may be
constructed from autonomous, time-reversal invariant system functions of the
form 
\begin{equation}
F(x_n,x_{n+1})=-\beta x_nx_{n+1}+\frac{_1}{^2}W(x_n)+\frac{_1}{^2}W\left(
x_{n+1}\right) ,
\end{equation}
where $\beta $ is a non-zero constant. The Cadzow's equation for this system
is 
\begin{equation}
x_{n+1}\stackunder{c}{=}\beta ^{-1}W^{\prime }\left( x_n\right) -x_{n-1},
\end{equation}
which has the merit of giving $x_{n+1}$ explicitly in terms of $x_n$ and $%
x_{n-1}$.

Now define the system amplitude to be given by 
\begin{equation}
U_n\left( x,y\right) =ke^{iF(x,y)/\hbar },
\end{equation}
where $k$ is some constant. Then from the unitarity condition $\left( \ref
{unit}\right) $ we find 
\begin{equation}
\left| k\right| ^2=\frac \beta {2\pi \hbar }.  \label{zz1}
\end{equation}
From $\left( \ref{comm}\right) $ we find 
\begin{equation}
\left[ \hat{x}_{n+1},\hat{x}_n\right] =\frac{-i\hbar }\beta ,  \label{zz2}
\end{equation}
so that the above co-ordinates for this system are normal. Moreover, with
the momentum $p_{n\text{ }}$ conjugate to $x_n$ defined by $\left( 10\right)
,$ we recover the conventional commutator 
\begin{equation}
\left[ \hat{p}_n,\hat{x}_n\right] =-i\hbar .
\end{equation}
This result is not expected to hold for systems which are not normal.

The operator equations of motion are found to be 
\begin{equation}
\hat{x}_{n+1}=\beta ^{-1}\hat{W}_n^{\prime }-\hat{x}_{n-1},
\end{equation}
where $\hat{W}^{\prime }$ is the diagonal operator 
\begin{equation}
\hat{W}_n^{\prime }\equiv \int dx\;|x,n\rangle \left\{ \frac{dW\left(
x\right) }{dx}\right\} \langle x,n|.
\end{equation}
From this we obtain the discrete time version of Ehrenfest's theorem; i.e. 
\begin{equation}
\langle \hat{x}_{n+1}\rangle =\beta ^{-1}\langle \hat{W}_n^{\prime }\rangle
-\langle \hat{x}_{n-1}\rangle 
\end{equation}
for expectation values over a physical state.

\subsection{Compatible operators}

In our theory the quantum dynamics is completely determined by the system
amplitude. Suppose now that the system is autonomous and time-reversal
invariant. This means that for each time $n$ we may write 
\begin{equation}
U_n\left( x,y\right) =U\left( x,y\right) ,
\end{equation}
where $U\left( x,y\right) $ is independent of $n$. For such a system there
may be constants of the motion comparable to the Maeda-Noether and Logan
invariants discussed in the classical theory.

Consider an operator $\hat{C}$ diagonal with respect to $\mathcal{B}^n$,
viz. 
\begin{equation}
\hat{C}\equiv \int dx\,|x,n\rangle C(x,\partial _x)\langle x,n|,
\end{equation}
where $C(x,\partial _x)$ is some differential operator of finite order.
Matrix elements of the commutator of $\hat{C}$ with $\hat{U}$ are given by 
\begin{equation}
\langle \phi |\left[ \hat{C},\hat{U}\right] |\psi \rangle =\int dxdy\,\phi
_n^{*}\left( x\right) \left\{ C(x,\partial _x)U\left( x,y\right) -\tilde{C}%
^{*}(y,\partial _y)U\left( x,y\right) \right\} \psi _n\left( y\right) ,
\end{equation}
where $|\psi \rangle $ and $|\phi \rangle $ are arbitrary physical states.
From this we arrive at the result

\textbf{Theorem: }\textit{A diagonal operator commutes with the timestep
operator of an autonomous system if }
\begin{equation}
C(x,\partial _x)U\left( x,y\right) =\tilde{C}^{*}(y,\partial _y)U\left(
x,y\right) .
\end{equation}

\textbf{Definition 5: }\textit{A diagonal operator which commutes with the
timestep operator of an autonomous system will be said to be compatible
(with the timestep operator).}

\textbf{Remark 10: }\textit{It is not necessary for a diagonal operator} $%
\hat{C}$ \textit{to be self-adjoint for it to be compatible with the
timestep operator.}

\textbf{Remark 11: }\textit{From the above we deduce that compatible
operators are invariants of the motion. To be explicit, consider a state} 
\TEXTsymbol{\vert}$\psi \rangle $ \textit{which is an eigenstate of the
diagonal operator} $\hat{C}$ \textit{with eigenvalue} $c$\textit{, i.e.} 
\begin{equation}
\hat{C}|\psi \rangle =c|\psi \rangle .
\end{equation}
\textit{Then we can show }
\begin{equation}
C(x,\partial _x)\psi _n\left( x\right) =c\psi _n\left( x\right) 
\end{equation}
\textit{and} 
\begin{equation}
C(x,\partial _x)\psi _{n+1}\left( x\right) =c\psi _{n+1}\left( x\right) .
\end{equation}

\textbf{Remark 12: }\textit{Given the system amplitude} $U\left( x,y\right) $
\textit{it may be very hard or perhaps even impossible to find any
compatible operators in closed form. It may be necessary to approximate such
an operator via a perturbative expansion, for example. This is the quantum
theory analogue of the problem of finding invariants of the motion for a
classical discrete time theory given some system function.}

\textbf{Remark 13:} \textit{Discrete time and continuous time quantum
mechanics pose dual problems in the following sense. In continuous time
quantum mechanics we are normally given a Hamiltonian and the problem is to
construct the time evolution operator. For a time independent Hamiltonian a
complete solution would require us to find all the eigenvalues} $E_\alpha $ 
\textit{and eigenstates} $|E_\alpha \rangle $ \textit{of} $\hat{H}$ \textit{%
and then use them in the formal solution} $\hat{U}_t=\exp \left\{ -i\hat{H}%
t/\hbar \right\} $ \textit{to write} 
\begin{equation}
\hat{U}_t=\sum_\alpha |E_\alpha \rangle e^{-iE_\alpha t/\hbar }\langle
E_\alpha |.
\end{equation}
\textit{This is in general a formidable problem. In discrete time quantum
mechanics the situation is the other way around. Given a system amplitude,
the problem is to find the compatible operators, if any. An important system
where answers can be found to all of these questions in both approaches is
the discrete time harmonic oscillator discussed in \S }$10$.

\section{Stroboscopic construction}

In principle it should always be possible to construct examples of discrete
time quantum systems by integrating the equation 
\begin{equation}
i\hbar \partial _t\hat{U}(t)=\hat{H}\hat{U}\left( t\right)   \label{evolve}
\end{equation}
for the evolution operator $\hat{U}(t)$ in continuous time quantum
mechanics, given the Hamiltonian $\hat{H}.$ The boundary condition 
\begin{equation}
\lim_{t\rightarrow 0}\hat{U}\left( t\right) =\hat{I}_{\mathcal{H}}
\end{equation}
ensures a unique solution. From this point of view the discussion outlined
above represents a stroboscopic approach, where the state vectors evolve
continuously but are looked at periodically in a non-destructive
(mathematical) sense. This examination of the state vector is \textbf{not}
the same as an observation collapsing the wavefunction.

For autonomous systems a formal solution to $\left( \ref{evolve}\right) $ is 
\begin{equation}
\hat{U}(t)\equiv \exp \left( i\hat{H}t/\hbar \right) .
\end{equation}
In this approach the transition amplitude $U(x,y;t)\equiv \langle
x,t|y,0\rangle $ corresponds to our system amplitude when $t=T$ and may be
evaluated in a number of ways. For example the Feynman path integral method
gives the formula 
\begin{equation}
\langle x,t|y,0\rangle \sim \int \left[ dz\right] \exp \left\{ \frac i\hbar
\int_0^tdt^{\prime }L\left( \dot{z},z,t^{\prime }\right) \right\} ,\;\;\;t>0,
\end{equation}
where $L$ is the Lagrangian, such amplitudes being functions of $t$ and the
end-points $x$ and $y$. The standard approach to the evaluation of such
integrals is, rather interestingly, based on the discretisation of time. The
time interval $\left[ 0,t\right] $ is partitioned into a finite number $N$
of equal steps, the integrand in the exponential is approximated suitably
(by what amounts to choosing a virtual path in our approach), the $N$
integrals evaluated, and then the limit $N\rightarrow \infty $ taken.

The relationship between this approach and our discrete time formalism
should now be clear, the basic difference being that we do not take the
limit $N\rightarrow \infty $. In a number of situations our system amplitude
will actually take the form 
\begin{equation}
U\left( x,y\right) \equiv \langle x,T|y,0\rangle \sim \exp \left\{ iF\left(
x,y\right) /\hbar \right\} ,  \label{point}
\end{equation}
and then for $N$ timesteps the transition amplitude $\langle x,NT|y,0\rangle 
$ becomes 
\begin{equation}
\langle x,NT|y,0\rangle \sim \int ...\int dx_1dx_2...dx_{N-1}\exp \left\{
iA^N/\hbar \right\} ,
\end{equation}
which emphasises the relationship further.

An important point which could be confusing is that the system function in (%
\ref{point}) does \textbf{not} correspond to a Logan constant for those
systems such as the harmonic oscillator where such an invariant can be
found. System functions in general are not expected to be invariants of the
motion.

An alternative method of constructing the transition amplitudes is to find
the Green's functions for the system. If the transition amplitude $U(x,y;t)$
satisfies the homogeneous Schr\"{o}dinger equation 
\begin{equation}
\left( i\hbar \partial _t-\overrightarrow{H_x}\right) U\left( x,y,t\right) =0
\end{equation}
with the boundary condition 
\begin{equation}
\lim_{t\rightarrow 0}U\left( x,y,t\right) =\delta ^3\left( \mathbf{x}\right)
,
\end{equation}
then the retarded and advanced Green's functions $G_R\left( x,y,t\right) $
and $G_A(x,y,t)$ are related to the transition amplitude by 
\begin{eqnarray}
G_R\left( x,y,t\right) &=&\theta \left( t\right) U\left( x,y,t\right) , 
\nonumber \\
G_A\left( x,y,t\right) &=&-\theta \left( -t\right) U\left( x,y,t\right) ,
\end{eqnarray}
and these satisfy the inhomogeneous equation 
\begin{equation}
\left( i\hbar \partial _t-\overrightarrow{H_x}\right) G\left( x,y,t\right)
=i\hbar \delta \left( t\right) \delta ^3\left( \mathbf{x}\right) .
\end{equation}
If we can solve this equation then we can immediately construct the
transition amplitude using the relation 
\begin{equation}
U\left( x,y,t\right) =G_R\left( x,y,t\right) -G_A\left( x,y,t\right) .
\end{equation}

\subsection{Example: the free Newtonian particle}

Given the Hamiltonian 
\begin{equation}
H=\frac{\mathbf{p.p}}{2m}
\end{equation}
in continuous time mechanics we can readily find the Greens functions in the
quantum theory. We find 
\begin{eqnarray}
G_R\left( \mathbf{x},\mathbf{y},t\right)  &=&\theta \left( t\right) \left( 
\frac{-im}{2\pi \hbar t}\right) ^{3/2}\exp \left\{ \frac i\hbar \frac{%
m\left( \mathbf{x-y}\right) .\left( \mathbf{x-y}\right) }{2t}\right\} , 
\nonumber \\
G_A\left( \mathbf{x},\mathbf{y},t\right)  &=&-\theta \left( -t\right) \left( 
\frac{-im}{2\pi \hbar t}\right) ^{3/2}\exp \left\{ \frac i\hbar \frac{%
m\left( \mathbf{x-y}\right) .\left( \mathbf{x-y}\right) }{2t}\right\} ,
\end{eqnarray}
from which we construct the transition amplitude 
\begin{equation}
U\left( \mathbf{x,y,}t\right) =\left( \frac{-im}{2\pi \hbar t}\right)
^{3/2}\exp \left\{ \frac i\hbar \frac{m\left( \mathbf{x-y}\right) .\left( 
\mathbf{x-y}\right) }{2t}\right\} ,\;\;\;\;\;t>0.
\end{equation}
This satisfies the closure condition $\left( \ref{unit}\right) $%
\begin{equation}
\int d^3\mathbf{y}U\left( \mathbf{x,y};T\right) U^{*}\left( \mathbf{z,y}%
;T\right) =\delta ^3\left( \mathbf{x-z}\right) 
\end{equation}
for a system amplitude and demonstrates the essential point that it is
possible to construct examples of discrete time quantum mechanics from
continuous time quantum mechanics. The converse need not be true. Given a
system amplitude, it may be impossible to find a compatible operator
equivalent to some second order Hamiltonian operator in continuous time
mechanics. It is not difficult to find examples of normal systems where this
occurs.

\section{The discrete time harmonic oscillator}

Given the discrete time harmonic oscillator system function 
\begin{equation}
F^n\equiv F\left( x_{n+1},x_n\right) =\frac{_1}{^2}\alpha \left(
x_{n+1}^2+x_n^2\right) -\beta x_{n+1}x_n
\end{equation}
we note that is an example of a normal system. This leads us to define the
system amplitude to be 
\begin{eqnarray}
U_n(x,y) &=&k\exp \left( iF(x,y)/\hbar \right)   \nonumber  \label{sys} \\
&=&k\exp \left( \frac i{2\hbar }\left[ \alpha x^2+\alpha y^2-2\beta
xy\right] \right) ,  \label{asys}
\end{eqnarray}
where $k$ is some constant. From the unitarity condition $\left( \ref{unit}%
\right) $ we find (\ref{zz1}) and from $\left( \ref{comm}\right) $ we find $%
\left( \ref{zz2}\right) $, so we see that the above system co-ordinates are
indeed normal. The self adjoint diagonal operator with operator component 
\begin{equation}
\overrightarrow{C_x}\equiv \frac{{}_1}{{}^2}\beta ^{-1}\left[ -\hbar
^2\partial _x^2+(\beta ^2-\alpha ^2)x^2\right]   \label{op2}
\end{equation}
is compatible with the system amplitude $\left( \ref{asys}\right) .$

The interpretation of this is that $\overrightarrow{C_x}$ is the operator
corresponding to the Logan constant for the classical discrete time harmonic
oscillator 
\begin{equation}
C=\frac{_1}{^2}\beta \left( x^2+y^2\right) -\alpha xy.
\end{equation}
To see this explicitly, consider the operators $\hat{x}_{n\text{ }}$and $%
\hat{x}_{n+1}.$ The Logan constant for the harmonic oscillator is quantised
according to the standard rule 
\begin{equation}
\hat{C}=\frac{_1}{^2}\beta \left( \hat{x}_n\hat{x}_n+\hat{x}_{n+1}\hat{x}%
_{n+1}\right) -\frac{_1}{^2}\alpha \left( \hat{x}_n\hat{x}_{n+1}+\hat{x}%
_{n+1}\hat{x}_n\right) .  \label{op1}
\end{equation}
A suitable co-ordinate representation of these operators with respect to the
basis $\mathcal{B}^{n\text{ }}$ is 
\begin{equation}
\hat{x}_{n\rightarrow }x,\;\;\;\;\hat{x}_{n+1}\rightarrow \eta x-i\frac
\hbar \beta \partial _x
\end{equation}
and then the operator $\left( \ref{op1}\right) $ is represented by $\left( 
\ref{op2}\right) .$

We see from the potential term in the differential operator $\left( \ref{op2}%
\right) $ that a complete set of physical states can be found as eigenstates
of the operator provided $\beta ^2>\alpha ^2.$ This corresponds precisely to
the elliptic region discussed in the classical theory.

If the constants satisfy the elliptic condition then we may construct
annihilation and creation operators for the system. These are diagonal with
respect to any of the bases $\mathcal{B}^{n\text{ }}$ and are given by 
\begin{eqnarray}
\hat{a}_n &\equiv &ie^{in\theta }\left[ \hat{x}_{n+1}-e^{i\theta }\hat{x}%
_n\right] =e^{in\theta }\int dx\,|x,n\rangle \left\{ \sqrt{1-\eta ^2}x+\frac
\hbar \beta \partial _x\right\} \langle x,n|,  \nonumber \\
\hat{a}_n^{+} &\equiv &-ie^{-i\theta }[\hat{x}_{n+1}-e^{-i\theta }\hat{x}%
_n]=e^{-i\theta }\int dx\,|x,n\rangle \left\{ \sqrt{1-\eta ^2}x-\frac \hbar
\beta \partial _x\right\} \langle x,n|.  \nonumber \\
&&
\end{eqnarray}
These operators satisfy the commutation relation 
\begin{equation}
\left[ \hat{a}_n,\hat{a}_n^{+}\right] =\frac{2\hbar \sqrt{1-\eta ^2}}\beta .
\end{equation}
Using the evolution relation (\ref{pos}) and the operator equation of motion 
\begin{equation}
\hat{x}_{n+1}=2\eta \hat{x}_n-\hat{x}_{n-1}  \label{heis}
\end{equation}
we find 
\begin{equation}
\hat{a}_{n+1}=\hat{a}_n,
\end{equation}
but this does not mean this operator is conserved. A conserved operator
according to our definition must be compatible with the timestep operator $%
\hat{U}_n.$ We find that the creation and annihilation operators satisfy the
relations 
\begin{eqnarray}
\hat{U}_n\hat{a}_n-e^{i\theta }\hat{a}_n\hat{U}_n &=&0,  \nonumber \\
\hat{U}_n\hat{a}_n^{+}-e^{-i\theta }\hat{a}_n^{+}\hat{U}_n &=&0,
\end{eqnarray}
which is reminiscent of various deformed commutators encountered in
q-mechanics. However, the operator ($\ref{op1})$ corresponding to the Logan
constant $C^n=\frac{{}_1}{{}^2}\beta a_n^{*}a_n$ is compatible with the
timestep operator and is therefore an invariant of the motion. We find 
\begin{eqnarray}
\hat{C} &=&\frac{_1}{^4}\beta \{\hat{a}^{+}\hat{a}+\hat{a}\hat{a}^{+}\}=%
\frac{{}_1}{{}^2}\beta \hat{a}^{+}\hat{a}+\frac{_1}{^2}\sqrt{1-\eta ^2}\hbar 
\nonumber \\
&=&\int dx\,|x,n\rangle \left\{ \frac{-\hbar ^2}{2\beta }\frac{\partial ^2}{%
\partial x^2}+\frac{{}_1}{{}^2}\beta \left( 1-\eta ^2\right) x^2\right\}
\langle x,n|=\int dx\,|x,n\rangle \overrightarrow{C_x}\langle x,n|. 
\nonumber \\
&&
\end{eqnarray}
This Logan invariant is a close analogue of the oscillator Hamiltonian in
continuous time mechanics and the eigenstates of the former follow the same
pattern as the eigenstates of the latter. For example, there is a ground
state \TEXTsymbol{\vert}$\Psi _0\rangle $ satisfying the relation 
\begin{equation}
\hat{a}_n|\Psi _0\rangle =0
\end{equation}
with normalisable wave function $\Psi _0\left( x\right) =\Psi _0\left(
0\right) \exp \left\{ -\frac{_1}{^2}\beta \sqrt{1-\eta ^2}x^2/\hbar \right\}
.$ This wave function is also an eigenstate of the Logan invariant operator,
with 
\begin{equation}
\overrightarrow{C}\Psi _0\left( x\right) =\frac{_1}{^2}\sqrt{1-\eta ^2}\hbar
\Psi _0\left( x\right) .
\end{equation}

These results hold only for $\left| \eta \right| <1$. We note from \S $5.3$
that $\beta =m(6+T^2\omega ^2)/6T$ and that 
\begin{equation}
\lim_{T\rightarrow 0}\frac{\overrightarrow{C}}T=-\frac{\hbar ^2}{2m}\partial
_x^2+\frac{_1}{^2}m\omega ^2
\end{equation}
when we identify our system with the Newtonian oscillator.

\section{The inhomogeneous oscillator}

In this section we discuss the inhomogeneous harmonic oscillator, which
serves as a prototype for the application of our quantisation principles to
field theories. We will use the source functional techniques of Schwinger to
obtain the ground state functional and various $n$-point functions of
interest. Because the Schwinger method deals with time ordered products we
should expect the discretisation of the time parameter to involve some
changes in the details of the calculations.

First, given a system function $F^n\equiv F\left( x_n,x_{n+1}\right) $ we
are free to introduce an external source in any convenient way, as
ultimately this will be set to zero. Our choice is to define the system
function $F^n\left[ j\right] $ in the presence of the external source as 
\begin{equation}
F^n\left[ j\right] \equiv F^n+\frac{_1}{^2}Tj_{n+1}x_{n+1}+\frac{_1}{^2}%
Tj_nx_n,
\end{equation}
a choice which allows the construction of time ordered product expectation
values directly. The action sum from time $MT$ to time $NT$ ($N>M)$ now
becomes 
\begin{equation}
A^{NM}\left[ j\right] =A^{NM}+\frac{_1}{^2}Tj_Mx_M+\frac{_1}{^2}%
Tj_Nx_N+T\sum_{n=M+1}^{N-1}j_nx_n,\;\;\;M<N
\end{equation}
from which Cadzow's equation of motion is found to be 
\begin{equation}
\frac \partial {\partial x_n}\left\{ F^n+F^{n-1}\right\} +j_n\stackunder{c}{=%
}0,\;\;\;M<n<N.
\end{equation}
Quantisation is introduced via the Schwinger action principle modified for
discrete time. We postulate that for an infinitesimal variation $\delta \hat{%
A}^{NM}\left[ j\right] $ of the action operator then 
\begin{equation}
\delta \langle \phi ,N|\psi ,M\rangle ^j=\frac i\hbar \langle \phi ,N|\delta 
\hat{A}^{NM}\left[ j\right] |\psi ,M\rangle ^j,\;\;\;M<N
\end{equation}
for any states $|\phi ,N\rangle $, $|\psi ,M\rangle $ at times $NT$, $MT$
respectively, with evolution in the presence of the source. Independent
variation of the $j_n$ for $M\leq n\leq N$ then leads to the equations 
\begin{eqnarray}
\frac{-i\hbar }T\frac \partial {\partial j_M}\langle \phi ,N|\psi ,M\rangle
^j &=&\frac{_1}{^2}\langle \phi ,N|\hat{x}_M|\psi ,M\rangle ^j  \nonumber \\
\frac{-i\hbar }T\frac \partial {\partial j_n}\langle \phi ,N|\psi ,M\rangle
^j &=&\langle \phi ,N|\hat{x}_n|\psi ,M\rangle ^j,\;\;\;\;\;\;\;\;\;\;M<n<N 
\nonumber \\
\frac{-i\hbar }T\frac \partial {\partial j_N}\langle \phi ,N|\psi ,M\rangle
^j &=&\frac{_1}{^2}\langle \phi ,N|\hat{x}_N|\psi ,M\rangle ^j.
\label{derivs}
\end{eqnarray}
Further application of the principle leads to expectation values of time
ordered product of operators, such as 
\begin{equation}
\left( \frac{-i\hbar }T\right) ^2\frac{\partial ^2}{\partial j_m\partial j_n}%
\langle \phi ,N|\psi ,M\rangle ^j=\langle \phi ,N|\widetilde{T}\ \hat{x}_m%
\hat{x}_n|\psi ,M\rangle ^j,\;\;\;M<m,n<N
\end{equation}
where the symbol $\widetilde{T}\ $denotes discrete time ordering. For
example, 
\begin{eqnarray}
\widetilde{T}\hat{x}_m\hat{x}_n &=&\left( \Theta _{m-n}+\frac{_1}{^2}\delta
_{m-n}\right) \hat{x}_m\hat{x}_n+\left( \Theta _{n-m}+\frac{_1}{^2}\delta
_{m-n}\right) \hat{x}_n\hat{x}_m  \nonumber \\
&=&\Theta _{m-n}\hat{x}_m\hat{x}_n+\delta _{m-n}\hat{x}_n\hat{x}_n+\Theta
_{n-m}\hat{x}_n\hat{x}_m,
\end{eqnarray}
where $\Theta _n$ is the discrete step function, defined by 
\begin{eqnarray}
\Theta _n &=&+1,\;\;n>0,\,\,\,\;\;\;  \nonumber \\
&=&0,\;\;\;\;n\leq 0
\end{eqnarray}
and $\delta _n$ is the Kronecker delta, defined by 
\begin{eqnarray}
\delta _n &=&+1,\;\;n=0,  \nonumber \\
&=&0,\;\;\;\;n\neq 0.
\end{eqnarray}

Given the harmonic oscillator system function 
\begin{equation}
F^n=\frac{m\left( x_{n+1}-x_n\right) ^2}{2T}-\frac{Tm\omega ^2}6\left(
x_{n+1}^2+x_{n+1}x_n+x_n^2\right) .
\end{equation}
then the classical discrete time harmonic oscillator in the presence of the
external source $j_{n\text{ }}$satisfies the equation 
\begin{equation}
x_{n+1}\stackunder{c}{=}2\eta x_n-x_{n-1}+\frac T\beta j_{n,}  \label{aa}
\end{equation}
where $\eta $ $=\alpha /\beta $ with 
\begin{equation}
\alpha =\frac{m(1-2T^2\omega ^2)}{6T},\;\;\;\beta =\frac{m(6+T^2\omega ^2)}{%
6T}.
\end{equation}
As discussed previously, elliptic (oscillatory) solutions occur for $\eta
^2<1$ whereas hyperbolic solutions occur for $\eta ^2>1.$ We will now
discuss these possibilities separately.

\subsection{The elliptic regime}

The importance of the elliptic regime $\eta ^2<1$ stems from the fact that
in field theory this corresponds to physical particle configurations of the
fields, i.e., solutions which can be normalised.

Now define the action of the $\left( \text{\textit{classical}}\right) $ 
\textit{discrete time displacement operator} $U_n$ by the rule 
\begin{equation}
U_nf_n\equiv f_{n+1}  \label{13}
\end{equation}
for any function of the index $n,\;$where $n$ is real. Then $\left( \ref{aa}%
\right) $ may be written in the form 
\begin{equation}
\left( U_n-2\eta +U_n^{-1}\right) x_n\stackunder{c}{=}\frac T\beta j_n.
\end{equation}

To solve $\left( \ref{aa}\right) $ for the elliptic case we first define the
following: since $\eta ^2<1\;$we write $c\equiv \cos (\theta )=\eta $ and $%
s\equiv \sin \left( \theta \right) =+\sqrt{1-\eta ^2}>0,$ taking $0<\theta
<\pi $. If we define$\;s_a\equiv \sin \left( a\theta \right) $ where $a$ is
real then a useful identity is 
\begin{equation}
s_as_{b-c}+s_bs_{c-a}+s_cs_{a-b}=0.  \label{id}
\end{equation}
From this we deduce 
\begin{equation}
s_{a+1}+s_{a-1}=2cs_a,
\end{equation}
which is equivalent to 
\begin{equation}
\left( U_a-2\eta +U_a^{-1}\right) s_a=0.
\end{equation}

We define the matrices 
\begin{equation}
\Lambda ^n=\frac 1s\left[ 
\begin{array}{cc}
s_{1+n} & -s_n \\ 
s_n & s_{1-n}
\end{array}
\right] ,\;\;\;\;\;\Lambda \equiv \Lambda ^1,
\end{equation}
and use $\left( \ref{id}\right) $ to prove 
\begin{equation}
\Lambda ^a\Lambda ^b=\Lambda ^{a+b}.
\end{equation}
If we write 
\begin{equation}
\mathbf{X}_n\equiv \left[ 
\begin{array}{c}
x_{n+1} \\ 
x_n
\end{array}
\right] ,\;\;\;\;\;\mathbf{J}_n\equiv \left[ 
\begin{array}{c}
\frac T\beta j_n \\ 
0
\end{array}
\right]
\end{equation}
then $\left( \ref{aa}\right) $ may be written in the form 
\begin{equation}
\mathbf{X}_n=\Lambda \mathbf{X}_{n-1}+\mathbf{J}_n.
\end{equation}
This equation may be readily solved using the properties of the $s_a$
functions and by diagonalising the matrix $\Lambda $. We choose Feynman
boundary conditions, specifying the particle to be at position $x_M$ in the
past (at time $MT$) and at position $x_N$ in the future (at time $NT$),
giving 
\begin{eqnarray}
&&ss_{N-M}x_n\stackunder{c}{=}ss_{N-n}x_M+ss_{n-M}x_N  \nonumber \\
&&\;\;\;\;\;\;\;\;\;\;\;+\frac T\beta \left\{
\sum_{m=M}^{n-1}s_{N-n}s_{M-m}j_m+s_{N-n}s_{M-n}j_n+%
\sum_{m=1+n}^Ns_{M-n}s_{N-m}j_m\right\} ,\;\;
\end{eqnarray}
$\;\;$which is valid only for $M<n<N.$ This can be tidied up into the form 
\begin{equation}
x_n=\frac{s_{N-n}x_M}{s_{N-M}}+\frac{s_{M-n}x_N}{s_{M-N}}-T%
\sum_{m=M}^NG_{NM}^{nm}j_m,
\end{equation}
where 
\begin{eqnarray}
G_{NM}^{nm} &=&-\frac{s_{N-n}s_{M-m}}{\beta ss_{_{N-M}}},\;\;\;\;\;M\leq
m<n<N  \nonumber \\
&=&-\frac{s_{N-n}s_{M-m}}{\beta ss_{_{N-M}}},\;\;\;\;\;\;\;M<m=n<N
\label{dprop} \\
&=&-\frac{s_{M-n}s_{N-m}}{\beta ss_{N-M}},\;\;\;\;\;\;M<n<m\leq N.  \nonumber
\end{eqnarray}
Then$\;G_{NM}^{nm}$ satisfies the inhomogeneous equation 
\begin{equation}
\beta \left( U_n-2\eta +U_n^{-1}\right) G_{NM}^{nm}=-\delta
_{n-m},\;\;\;\;M<n<N.
\end{equation}

Up to this stage we have taken $N>M$ with both finite, but normally we will
be interested in the scattering limit $N\rightarrow +\infty ,\;M\rightarrow
-\infty .$ Also, we have appeared to have overlooked the possibility that $%
s_{_{N-M}}$ vanishes in the denominator of the propagator $\left( \ref{dprop}%
\right) $ for some values of $N$ and $M$. We shall address both of these
issues directly now.

Our method of avoiding possible singularities is to extend the Feynman $%
-i\epsilon $ prescription to the $\theta $ parameter. By inspection of the
equation 
\begin{equation}
\eta \equiv \cos \left( \theta \right) =\frac{6-2T^2\omega ^2}{6+T^2\omega ^2%
}  \label{eta}
\end{equation}
we deduce that 
\begin{equation}
\omega ^2\rightarrow \omega ^2-i\epsilon \Rightarrow \theta \rightarrow
\theta -i\epsilon ,\;\;\;\eta \rightarrow \eta +i\epsilon .
\end{equation}
With this deformation of the $\theta $ parameter and taking the limit $%
N\rightarrow +\infty ,\;M\rightarrow -\infty ,$ we find 
\begin{equation}
x_n=\tilde{x}_n-T\sum_{m=-\infty }^\infty G_F^{n-m}j_m
\end{equation}
where $\tilde{x}_n$ satisfies the homogeneous equation 
\begin{equation}
\left( U_n-2\eta +U_n^{-1}\right) \tilde{x}_n=0
\end{equation}
and 
\begin{equation}
G_F^{n-m}=\frac 1{2\beta is}\left( e^{i\left( m-n\right) \theta }\Theta
_{n-m}+\delta _{m-n}+e^{i\left( n-m\right) \theta }\Theta _{m-n}\right) .
\end{equation}
This is the discrete time analogue of the harmonic oscillator Feynman
propagator and reduces to it in the limit $T\rightarrow 0,\;nT\rightarrow t$%
. A direct application of the discrete time Schwinger action principle to
the operator equation of motion 
\begin{equation}
\left( U_n-2\eta +U_n^{-1}\right) \hat{x}_n=\frac T\beta j_n
\end{equation}
then gives the ground state vacuum functional 
\begin{equation}
Z\left[ j\right] =Z\left[ 0\right] \exp \left\{ \frac{-iT^2}{2\hbar }%
\sum_{n,m=-\infty }^\infty j_nG_F^{n-m}j_m\right\} ,
\end{equation}
essentially solving the quantum problem. Using this result and $\left( \ref
{derivs}\right) $ we find that in the limit $j\rightarrow 0$%
\begin{equation}
\langle 0|\hat{x}_n\hat{x}_n|0\rangle =\frac \hbar {2\beta
s},\;\;\;\;\langle 0|\hat{x}_{n+1}\hat{x}_n|0\rangle =\frac \hbar {2\beta
s}e^{-i\theta }.  \label{resx}
\end{equation}
From this we deduce 
\begin{equation}
\langle 0|[\hat{x}_{n+1},\hat{x}_n]|0\rangle =\frac{-i\hbar }\beta ,
\end{equation}
which agrees exactly with the discrete time oscillator commutation relation 
\begin{equation}
\lbrack \hat{x}_{n+1},\hat{x}_n]=\frac{-i\hbar }\beta  \label{com1}
\end{equation}
found previously.

Further ground state expectation values of commutators may be obtained by
using the result 
\begin{equation}
\langle 0|\tilde{T}\hat{x}_m\hat{x}_n|0\rangle =i\hbar G_F^{n-m}=\frac \hbar
{2\beta \sin \theta }e^{-i|n-m|\theta }.
\end{equation}
For example, we find 
\begin{equation}
\langle 0|\left[ \hat{x}_{n+2},\hat{x}_n\right] |0\rangle =\frac{-2i\hbar
\eta }\beta ,
\end{equation}
which agrees with the commutator 
\begin{equation}
\left[ \hat{x}_{n+2},\hat{x}_n\right] =\frac{-2i\hbar \eta }\beta
\end{equation}
obtained from the operator equation of motion $\left( \ref{heis}\right) $
and the commutator $\left( \ref{com1}\right) .$

Another verification of the consistency of our methods is that we may use $%
\left( \ref{resx}\right) $ directly to find the ground state expectation
value of the Logan invariant (\ref{op1}) for the discrete time harmonic
oscillator. We find 
\begin{equation}
\;\langle 0|\hat{C}^n|0\rangle =\frac{_1}{^2}\hbar \sqrt{1-\eta ^2},
\end{equation}
which agrees exactly with previous results.

\subsection{The hyperbolic regime}

Because $T\omega $ is real and positive the controlling parameter $\eta $ as
given by (\ref{eta}) takes vales only in the regions 
\begin{eqnarray}
\text{elliptic} &:&\text{ \ }-1<\eta <1:\;\;\;0<T\omega <2\sqrt{3}  \nonumber
\\
\text{parabolic} &:&\;\;\;\;\;\;\;\;\;\eta =-1:\;\;\;T\omega =2\sqrt{3} 
\nonumber \\
\text{hyperbolic} &:&\;-\infty <\eta <-1:\;\;2\sqrt{3}<T\omega .
\label{range}
\end{eqnarray}
If we parametrise $\eta $ by the rule $\eta =\cos \left( z\right) $ where $z$
is complex then if we take 
\begin{eqnarray}
\eta  &=&\cos \theta :\;\;\;\;0<T\omega <2\sqrt{3}  \nonumber \\
\eta  &=&-\cosh \lambda :\;2\sqrt{3}<T\omega ,
\end{eqnarray}
then the range of possibilities (\ref{range}) corresponds to a contour $%
\Gamma $ in the complex $z=\theta -i\lambda $ plane which runs just below
the real axis from the origin to $\pi $ and then runs from $\pi $ to $\pi
-i\infty .$ The elliptic regime corresponds to values of $z$ on the first
part of the contour, for which $\lambda =0+\epsilon ,$ where $\epsilon $ is
infinitesimal and positive, corresponding to the Feynman $-i\epsilon $
prescription.

The hyperbolic region corresponds to the part of the contour given by \ \ $%
z~=~\pi -i\lambda ,:\lambda >0.$ For this region analytic continuation of
the $s_n$ functions leads to 
\begin{equation}
s_n\rightarrow i\left( -1\right) ^{n+1}\tilde{s}_n
\end{equation}
where $\tilde{s}_n\equiv \sinh \left( n\lambda \right) .$ From this the
analytic continuation of the finite interval propagator (\ref{dprop}) gives 
\begin{eqnarray}
\tilde{G}_{NM}^{nm} &\equiv &\frac{-\left( -\right) ^{n-m}}{\beta \tilde{s}%
\tilde{s}_{N-M}}\left\{ \tilde{s}_{N-n}\tilde{s}_{M-m}\Theta _{n-m}+\tilde{s}%
_{N-m}\tilde{s}_{M-m}\delta _{n-m}+\Theta _{m-n}\tilde{s}_{M-n}\tilde{s}%
_{N-m}\right\} ,\;  \nonumber \\
\;
&&\;\;\;\;\;\;\;\;\;\;\;\;\;\;\;\;\;\;\;\;\;\;\;\;\;\;\;\;\;\;\;\;\;\;\;\;%
\;M<n,m<N,
\end{eqnarray}
which satisfies the equation 
\begin{equation}
\beta \left\{ U_n-2\eta +U_n^{-1}\right\} \tilde{G}_{NM}^{nm}=-\delta _{n-m}.
\end{equation}
Taking the limit $N=-M\rightarrow \infty $ gives the infinite interval
propagator 
\begin{equation}
\tilde{G}_F^{n-m}\equiv \frac{\left( -\right) ^{1+n-m}}{2\beta \tilde{s}}%
\left\{ e^{\left( m-n\right) \lambda }\Theta _{n-m}+\delta _{n-m}+e^{\left(
n-m\right) \lambda }\Theta _{m-n}\right\}
\end{equation}
which satisfies the equation 
\begin{equation}
\beta \left\{ U_n-2\eta +U_n^{-1}\right\} \tilde{G}_F^{n-m}=-\delta _{n-m}.
\end{equation}

\subsection{Comment}

The elliptic and hyperbolic Feynman boundary condition propagators can be
summarised in the analytic form 
\begin{eqnarray}
\Delta _F^n\left( \eta \right)  &=&\frac{T\left( 2+\cos z\right) }{6mi\sin z}%
e^{-i|n|z}  \nonumber \\
&=&\frac{T\left( 2+\cos z\right) }{6mi\sin z}\left( e^{-inz}\Theta _n+\delta
_n+e^{inz}\Theta _{-n}\right) ,
\end{eqnarray}
where $\eta =\cos z$ and $z$ lies somewhere on the contour $\Gamma $
discussed above. Then $\Delta _F^n\left( \eta \right) $ satisfies the
equation 
\begin{equation}
\beta \left\{ U_n-2\eta +U_n^{-1}\right\} \Delta _F^n\left( \eta \right)
=-\delta _n.
\end{equation}

However, physical states correspond only to points in the elliptic regime,
as the wave functions will not remain normalisable in time otherwise. For
example, we showed previously that the ground state wavefunction for the
discrete time oscillator in the elliptic regime is given by 
\begin{equation}
\Psi _0\left( x\right) =\Psi _0\left( 0\right) \exp \left( -\frac{_1}{^2}%
\beta \sqrt{1-\eta ^2}x^2/\hbar \right)
\end{equation}
demonstrating that analytic continuation to the hyperbolic regime will not
give a normalisable ground state wave-function.\ 

\section{Concluding remarks}

In this paper we have shown that a consistent approach to the discretisation
of time results in a consistent dynamical framework. Once the initial
psychological hurdle of accepting a dynamics without time derivatives has
been jumped, then such a theory becomes as reasonable as continuous time
dynamics. Indeed, by taking the fundamental time interval $T$ small enough,
it would appear possible to duplicate or approximate conventional theory as
closely as required. The obvious question, \textit{why consider discrete
time mechanics at all?} has two replies. First, it may be the case after all
that there is some fundamental limit to time intervals, and so it becomes a
matter of curiosity as to how far we can go along that road. Second, there
may be some novel properties in this approach which could prevent or
alleviate the notorious problems with divergences which plague conventional
field theories. It is worth considering any approach to the regularisation
in field theory which is based on just one assumption, namely that of a
discrete time. The behaviour of the discrete time oscillator holds the
promise of potentially useful properties which may provide a cutoff for
particle energy.

In the next paper we apply our methods to classical field theories. However
no divergence problems appear at that stage. In the third paper of this
series, we shall consider second quantisation, and the issues to do with
divergences of Feynman diagrams will be discussed in subsequent papers in
some detail.

\section{Acknowledgements}

We are grateful to Prof. J Lukierski for providing us with important
references and discussion on discrete time mechanics.

\newpage\

\end{document}